\documentclass[aps,twocolumn,showpacs,preprintnumbers,amsmath,amssymb,nofootinbib,superscriptaddress,showkeys,floatfix]{revtex4}

\usepackage{epsfig}
\usepackage{graphicx}

\def\slashchar#1{\setbox0=\hbox{$#1$}
   \dimen0=\wd0 \setbox1=\hbox{/} \dimen1=\wd1
   \ifdim\dimen0>\dimen1 \rlap{\hbox to \dimen0{\hfil/\hfil}} #1
   \else  \rlap{\hbox to \dimen1{\hfil$#1$\hfil}} / \fi}

\newcommand{\lrd}{\raisebox{0.09em}{$\stackrel{\scriptstyle\leftharpoonup\hspace{-.7em}\rightharpoonup}{D} $}}

\begin{document}

\title{Scalar-isoscalar states in the large-$N_c $ Regge approach}

\author{Enrique Ruiz Arriola}
\email{earriola@ugr.es}
\affiliation{Departamento de
  F\'{\i}sica At\'omica, Molecular y Nuclear, Universidad de Granada, E-18071 Granada, Spain.}
\author{Wojciech Broniowski} 
\email{Wojciech.Broniowski@ifj.edu.pl}
\affiliation{The H. Niewodnicza\'nski Institute of Nuclear Physics, Polish Academy of Sciences, PL-31342 Krak\'ow, Poland}
\affiliation{Institute of Physics, Jan Kochanowski University, PL-25406~Kielce, Poland}

\date{\today}

\begin{abstract} 
Scalar-isoscalar states ($J^{PC}=0^{++}$) are investigated within the
large-$N_c$ Regge approach. We elaborate on the consequences of
including the lightest $f_0(600)$ scalar-isoscalar state into such an
analysis, where the position of $f_0(600)$ fits very well into the
pattern of the radial Regge trajectory.  Furthermore, we point out
that the pion and nucleon spin-0 gravitational form factors, recently
measured on the lattice, provide valuable information on the low-mass
spectrum of the scalar-isoscalar states on the basis of the
scalar-meson dominance in the spin-0 channel.  Through the fits to
these data we find $m_\sigma= 450-600$~MeV.  We compare the
predictions of various fits and methods.  An analysis of the QCD
condensates in the two-point correlators provides further constraints
on the parameters of the scalar-isoscalar sector.  We find that a
simple two-state model suggests a meson nature of $f_0(600)$, and a
glueball nature of $f_0(980)$, which naturally explains the ratios of
various coupling constants.  Finally, we note that the fine-tuned
condition of the vanishing dimension-2 condensate in the Regge
approach with infinitely many scalar-isoscalar states 
yields a reasonable value for the mass of the lighest glueball state. 
\end{abstract}



\maketitle

\section{Introduction \label{sec:intro}}

The history and status of the $\sigma$-meson has been quite
vacillating (for reviews see e.g. \cite{Klempt:2007cp,2008AIPC.1030.....R,Achasov:2009ee}
and references therein). A scalar-isoscalar state with a mass of $
\sim 500~{\rm MeV}$ was originally proposed in the
fifties~\cite{PhysRev.98.783} as an ingredient of the nucleon-nucleon
force providing saturation and binding in nuclei. Along the years,
there has always been some arbitrariness in the ``effective'' or
``fictitious'' $\sigma$ meson mass and the coupling constant to the
nucleon, partly due to the lack of other sources of information. For
instance, in the very successful Charge Dependent (CD) Bonn
NN-potential~\cite{Machleidt:2000ge}, any partial wave
$^{2S+1}L_J$-channel is fitted with a different scalar-isoscalar meson
mass and coupling. The $\sigma$-meson was also introduced as the
chiral partner of the pion to account for spontaneous breaking of the
chiral symmetry~\cite{GellMann:1960np}. The lack of confidence in its
existence motivated taking its mass to infinity, yielding the
non-linear sigma model~\cite{Weinberg:1968de}, which is the modern
starting point for the Chiral Perturbation Theory~\cite{Gasser:1983yg}.

During the last decade, the situation has steadily changed, and the
$\sigma$-meson has been finally resurrected~\cite{Tornqvist:1995ay},
culminating with the inclusion of the $0^{++}$ resonance in the
Particle Data Group review (PDG)~\cite{Yao:2006px} as the $f_0 (600)$
state, seen as a $\pi\pi$ resonance. It has wide-spread values for the
mass, $400-1200~{\rm MeV}$, and for the width, $600-1200~{\rm
  MeV}$~\cite{vanBeveren:2002mc}. A rigorous definition of the
$\sigma$ as a $\pi\pi$ resonance requires that it be a pole of the
$\pi\pi$ scattering amplitude in the $(J,T)=(0,0)$ channel in the
second Riemann sheet in the Mandelstam variable $s$. Within such a
framework, the uncertainties have recently been narrowly sharpened
with a benchmark determination based on the Roy equations with
constraints from the chiral symmetry~\cite{Caprini:2005zr}, yielding
the value~\footnote{We use this definition for the pole mass in the
 $\sqrt{s}$ variable. A better one is $s_\sigma = m_\sigma^2 - i
  \Gamma_\sigma m_\sigma$, which coincides with the previous one in
  the narrow resonance limit.}
\begin{eqnarray}
m_\sigma - i \Gamma_\sigma /2 = 441^{+16}_{-8} - i 272^{+9}_{-12} {\rm ~MeV}. \label{444}
\end{eqnarray}
The analysis of Ref.~\cite{Kaminski:2006qe} yields a result with
somewhat higher $m_\sigma$, $473\pm 6 \pm 11 -i 257 \pm 5 \pm 2$~MeV
(with the errors statistical and systematic, respectively), while the
unitarized Chiral Perturbation Theory ($\chi$PT) gives a bit lower
value of the mass, $401^{+12}_{-16} - i
277^{+23}_{-26}$~MeV~\cite{Nieves:2009ez}. Nevertheless, various
determinations of the pole mass agree with the values (\ref{444})
within the uncertainties. These accurate determinations make somewhat
tricky the original question on {\em what} $\sigma$ mass should be
used {\it a priori} within a meson-exchange picture, due to the very
large width of the resonance.  Moreover, the determinations mentioned
above do not imply necessarily the standard assignment of the linear
sigma-model where one takes $(\sigma, \vec \pi)$ as chiral partners in
the $(\frac12,\frac12) $ representation of the chiral $SU(2)_R \otimes
SU(2)_L$ group. A priori, the $\sigma$ state could also belong to the
$(0,0)$ representation. Admittedly, the heated debate on the nature of
the $\sigma$ meson is not completely over, with the
tetraquark~\cite{Hooft:2008we} (see also \cite{Fariborz:2009cq}) or
glueball interpretations~\cite{Kaminski:2009qg} considered (see, e.g,
Ref.~\cite{Klempt:2007cp} and numerous references therein).

Our goal in the present paper is to point out that the large-$N_c$
Regge models may provide valuable insight into this problem. Clearly,
the numbers in Eq.~(\ref{444}) represent the values for $N_c=3$. We
recall that in the large-$N_c$ limit QCD can be mapped onto a theory
of {\it infinitely} many stable and non-interacting mesons and
glueballs~\cite{'tHooft:1973jz,Witten:1979kh}. In the standard large-$N_c$
counting, meson and glueball masses are both ${\cal O} (N_c^0)$
whereas their widths are ${\cal O} (N_c^{-1})$ and ${\cal O}
(N_c^{-2})$, respectively. However, there is also a
large-$N_c$-suppressed mass shift which makes a literal use of
Eq.~(\ref{444}) somewhat questionable. Actually, lattice calculations
have shown that QCD$_\infty$ is generally not too far from QCD at
$N_c=3$~(see, e.g., Ref.~\cite{Teper:2008yi} and references therein),
at least for certain observables. The lightest scalar-isoscalar state
(commonly denoted by $\sigma$), due to its unusually large width,
seems to be an exceptional case, and in this paper we try to address
its nature and phenomenological consequences within the framework of
the large-$N_c$ Regge models.

Large-$N_c$-motivated investigations for the $\pi\pi$ or NN
interactions keep only the lowest $0^{++}$ state, and hence constitute
the low-energy analyses (see e.g. Ref.~\cite{Sannino:1995ik} for an
early investigation). The studies of the $\pi\pi$ scattering, based on a
large-$N_c$ scaling of the $N_c=3$ parameters and unitarized via the
Inverse Amplitude Method (IAM) applied to the $\chi$PT amplitudes,
provide an $N_c$ dependence of resonances which, regarding the
$\sigma$ state, depends on the details of the scheme used. While the
one-loop coupled channel approach~\cite{Pelaez:2003dy} yields a very
wide range of $m_\sigma$ and a large width (in an apparent
contradiction with the standard large-$N_c$
counting~\cite{'tHooft:1973jz,Witten:1979kh} when extrapolated to $N_c
\gg 10$), the two-loop approach~\cite{Pelaez:2006nj} produces a large
shift (by a factor of $2$) of $m_\sigma$ when going from $N_c=3$ to
$N_c = \infty$ (the corresponding mass shift is small in the case of
the $\rho$ meson).  One should note the large uncertainties of the
two-loop IAM, documented in Ref.~\cite{Nieves:2001de}.  Along similar
large-$N_c$ counting, the $K$-matrix approach~\cite{Harada:2003em} gives
$m_\sigma \sim 2 \sqrt{\pi} f_\pi = {\cal O} (\sqrt{N_c})$. The
Bethe-Salpeter approach~\cite{Nieves:1999bx} yields an estimate
$m_\sigma \sim 500 {\rm ~MeV}$~\cite{CalleCordon:2008eu} at $N_c \to
\infty$. A more refined analysis using the large-$N_c$ consistency
conditions between the unitarization and resonance saturation suggests
$m_\rho - m_\sigma = {\cal O}(N_c^{-1})$~\cite{Nieves:2009ez}. 

The large-$N_c$ analysis of the NN force was carried out in
Refs.~\cite{Kaplan:1995yg,Kaplan:1996rk} at the quark level and in
Ref.~\cite{Banerjee:2001js} at the hadronic level, providing a
justification of the meson-exchange picture. Moreover, it indicates
{\it what} mesons are the leading ones, namely the infinite tower of
$\sigma$, $\pi$, $\rho$, $\omega$, $A_1$, {\em etc.}, states. Meson widths
enter as relative $1/N_c^2$ corrections to the potential, on equal
footing with many other effects (the spin-orbit force, relativistic
dynamics, or the inclusion of other mesons), independently on how
large the $\sigma$ width is in the real $N_c=3$ world. In this regard
the large-$N_c$ analysis may provide a reliable determination of the
asymptotic value of the $\sigma$ mass at $N_c \to \infty$. A fit to
the $^1S_0$ leading-$N_c$ potential yields 
$m_\sigma = 501(20) {\rm~MeV}$~\cite{CalleCordon:2008eu}.  Moreover, it was shown in
Ref.~\cite{CalleCordon:2009ps} within the large-$N_c$ framework, that
the correlated $2\pi$ exchange corresponds to a Yukawa potential where
the corresponding Yukawa mass differs by a $1/N_c$ correction to the
pole mass or equivalently to the Breit-Wigner mass (the difference
between both masses is ${\cal O} (N_c^{-2})$~\cite{Nieves:2009kh}).
Finally, we should also note that using an alternative large-$N_c$ 
counting~\cite{Corrigan:1979xf,Kiritsis:1989ge}, the scalar meson
behaves as a $\bar q q \bar q q$ state~\cite{Sannino:2007yp} and its
width is ${\cal O}(N_c^{-2})$.

In the present work we take into account an infinite number of
scalar-isoscalar states. In order to gather information on their
coupling to hadrons and the vacuum, we use a Regge model for the
radial trajectories and study correlation functions between the
energy-momentum tensor. Surprisingly, we find in
Section~\ref{sec:traj} that quite naturally {\it all} the
scalar-isoscalar states can be described by a single radial Regge
trajectory with half the standard slope. The mass of the $\sigma$
state can be deduced from this trajectory as the mass of the lowest
state.  As a consequence, there seems to be no obvious difference
between mesons and glueballs, as far as the spectrum is concerned.
Despite this surprising ordering with a sort of a radial quantum number,
we should warn the reader that states of different nature may not necessarily be
characterized by a single quantum number and other interpretations 
are considered in the literature. For example,
$f_0(980)$ is often regarded as $\bar q q \bar qq$, $f_0(1370)$ as mainly $\bar
uu+\bar dd$, $f_0(1500)$ as mainly a glueball, and $f_0(1710)$ dominantly as
$\bar ss$~\cite{Klempt:2007cp}. In such a case, there is no
trajectory revealing the sigma meson mass, and the observed
regularity would be accidental.

Existing lattice studies of hadronic matrix elements of the energy-momentum tensor, 
the so-called gravitational form factors of the pion
and nucleon, provide a very valuable complementary information on the
$\sigma$ properties, with the large-$N_c$ interpretation behind
(Section~\ref{sec:grav}). Unfortunately, the data are too noisy as to
pin down the coupling of the excited scalar-isoscalar states to the
energy-momentum tensor. Nevertheless, useful information confirming
the mass estimates for the $\sigma$-meson can be extracted
(Sect.~\ref{sec:grav}).  It is observed from the lattice data that the
$\sigma$ mass grows with the value of the light current quark mass in
a natural way.

It is interesting to discern the nature of the $\sigma$ state from an
analysis of a truncated spectrum. The minimum number of states,
allowed by certain sum rules and low energy theorems, is just two. In
Section~\ref{sec:more_states} we undertake such an analysis, which
suggests that $f_0(600)$ is a $\bar q q $ meson, while $f_0(980)$ is a
glueball.  Finally, in Section~\ref{sec:dim2} we address the rather
fundamental issue of the existence of dimension-2 condensates within
the context of the Operator Product Expansion of QCD. Such objects
appear naturally in the Regge approach and could not vanish if only a
finite number of states were kept. Actually, we show that the
dimension-2 condensate may vanish when infinitely many states are
considered, in which case a fine-tuning condition on the $\sigma$ mass
is needed. Finally, in Section~\ref{sec:concl} we come to our
conclusions.

\section{Scalar-isoscalar Regge trajectories \label{sec:traj}}

\begin{figure}[tb]
\includegraphics[width=.47\textwidth]{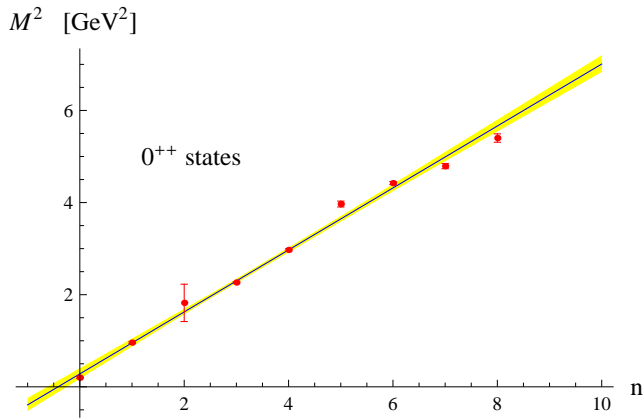}
\caption{(Color online) Radial Regge trajectory corresponding to the squared mass of all
$J^{PC}=0^{++}$ scalar-isoscalar states listed in the
PDG tables~\cite{Amsler:2008zz}.  The four heaviest $0^{++}$ states are not
yet well established and are omitted from the PDG summary table. 
The error bars correspond to the errors in the determination of the 
square of mass, listed in~\cite{Amsler:2008zz}. The straight line is the result of Fit~I with {\em half} the standard Regge slope,
according to Eq.~(\ref{eq:massS}), with $a$ and $m_\sigma$ from Eq.~(\ref{Regge8fit}). The band reflects the uncertainty 
in the parameters of Eq.~(\ref{Regge8fit}).}
\label{fig:regge-trajectory}
\end{figure}

In this section the mass of the $\sigma$-meson is determined by
extrapolating the upper part of the $J^{PC}=0^{++}$ spectrum in the
large-$N_c$ Regge approach. In Ref.~\cite{Anisovich:2000kxa} the
analysis of the radial and rotational Regge trajectories was carried
out. In Ref.~\cite{Anisovich:2005jn} the scalar sector was studied in
a greater detail, including the modeling of the meson widths. Two
parallel radial trajectories could then be identified, including three
states per trajectory, which has resulted in rather poor statistics.
In Fig.~\ref{fig:regge-trajectory} we show the squared mass of all
nine $J^{PC}=0^{++}$ isoscalar states listed in the PDG
tables~\cite{Amsler:2008zz}, given for completeness in
Table~\ref{tab:regge-fits}. Note that the four heaviest $0^{++}$ states
are omitted from the PDG summary table and in principle need
confirmation.  As we can see, an approximate straight line behavior
looks quite appealing.  To improve the statistics, we propose to fit
with {\em half} the standard slope, thus including all the PDG
scalar-isoscalar $0^{++}$ states together in the mass formula
\begin{eqnarray}
M_S (n)^2 = \frac{a}{2} n + m_\sigma^2, \label{eq:massS}
\end{eqnarray} 
where $a=2 \pi \sigma$, and $\sigma$ is the string tension.
The same strategy was recently applied in Ref.~\cite{dePaula:2009za}.

In the first method, called Fit~I, we minimize the $\chi^2$ variable  
\begin{eqnarray}
\chi^2 = \sum_n \left(\frac{M_{f,n}-M_S(n)}{\Delta M_{f,n}} \right)^2 .
\end{eqnarray}
We get, after {\em excluding} the lowest $n=0$ state and using the 8 remaining PDG states, the 
central parameter 
values $a=1.35{\rm ~GeV}^2$ and $m_\sigma=530~{\rm MeV}$, with $\chi^2/{\rm DOF}=\chi^2/(8-2)=9.9$. 

A non-trivial issue, not addressed in the analyses of
Refs.~\cite{Anisovich:2000kxa,Anisovich:2005jn}, is the determination of 
errors of the parameters in the Regge fits. Actually, the large value of  
$\chi^2/{\rm DOF} \sim 10 \gg 1 $ found above
prevents a reliable error determination of the model parameters $a$
and $m_\sigma$ and calls for corrections from different
sources. A common way to overcome this difficulty is to rescale the
weights in the $\chi^2$ variable. In our case we simply replace $\Delta M_{f,n} \to 3 \Delta M_{f,n}$, such that 
$\chi^2/{\rm DOF} \sim 1$. In this case we get 
\begin{eqnarray}
a=1.35(5) {\rm ~GeV}^2, \;\;\; m_\sigma = 530(95) {\rm ~MeV} \;\;\; ({\rm Fit~I}). \nonumber \\
\label{Regge8fit}
\end{eqnarray}

In the usual approaches (see, e.g.,
Refs.~\cite{Anisovich:2005jn,Surovtsev:2008xr}) a difference between the
bare and physical pole masses, based on the $K$-matrix or the Breit-Wigner
formula, is made. This generally introduces some model dependence and
thus there may be systematic uncertainties in the masses related to
the precise definition of the resonance parameters.

As we can see from Eq.~(\ref{Regge8fit}), the fit is quite acceptable and implies 
the value of the string tension
$\sqrt{\sigma} =463(9)~{\rm MeV}$. We note that the phenomenological value is
$\sqrt{\sigma} = 420~{\rm MeV}$ from a global fit to the excited meson
spectrum~\cite{Anisovich:2000kxa}. 

Formula (\ref{eq:massS}) is equivalent
to two parallel radial Regge trajectories with the {\em standard} slope
\begin{eqnarray}
M_{S,-} (n)^2 &=& a \, n + m_\sigma^2,  \nonumber  \\
M_{S,+} (n)^2 &=& a \, n + m_\sigma^2 + \frac{a}{2}. \label{twotraj}
\end{eqnarray} 
From here we get the mass-splitting formula 
\begin{eqnarray}
M_{S,+} (n)^2 -M_{S,-} (n)^2 = \frac{a}{2}=\pi \sigma  
\end{eqnarray} 
and, in particular,
\begin{eqnarray}
2(m_{f_0(980)}^2 - m_\sigma^2) = a = 2\pi \sigma,  
\end{eqnarray} 
which works extremely well for the value $m_{f_0} =980 {\rm ~MeV}$ and
$m_\sigma=530 {\rm ~MeV}$, yielding $1.36 {\rm ~GeV}^2 $ for the l.h.s.
vs. $ 1.35 {\rm ~GeV}^2$ for the r.h.s. 

Given the fact that all states fit quite naturally into the pattern, we see no obvious way
how glueball states could be singled out solely on the
basis of belonging to one of the trajectories (\ref{twotraj}). Also note that a glueball corresponds to a bound state in
gluodynamics, or equivalently QCD for infinitely heavy quarks (zero
active flavors), while the states obtained here correspond to the
limit of light quarks (two or three active flavors).  

\begin{table}[tb]
\begin{tabular}{|c|c|c|c|c|c|}
\hline 
Resonance  & $M$  [MeV]    & $\Gamma$ [MeV] & $n$ & $M$ (Fit I) & $M$ (Fit II) \\ 
\hline 
$f_0(600)$    & $400-1200$      & $500-1000$    & 0 & $530$ & $556$       \\ 
$f_0(980)$    & $980(10)$      & $70(30)$ & 1 & $976$ & $983$    \\ 
$f_0(1370)$   & $1350(150) $      & $400(100)$  & 2 & $1275$ & $1274$        \\ 
$f_0(1500)$   & $1505(6) $      & $109(7)$   & 3 & $1516$ & $1510$      \\ 
$f_0(1710)$   & $1724(7) $      & $137(8)$   & 4 & $1724$  &$1714$     \\ 
$f_0(2020)$   & $1992(16) $      & $442(60)$ & 5 & $1909$  &$1896$       \\ 
$f_0(2100)$   & $2103(8) $      & $209(19)$  & 6 & $2078$  &$2062$      \\ 
$f_0(2200)$   & $2189(13) $      & $238(50)$ & 7 & $2234$  &$2215$       \\ 
$f_0(2330)$   & $2321(30) $      &  $223(30)$ & 8 & $2380$ & $2359$         \\ 
\hline
\end{tabular}
\caption{\label{tab:regge-fits} PDG values of resonance
  parameters~\cite{Amsler:2008zz}, compared to the fits to the radial
  Regge spectrum of the $0^{++}$ scalar-isoscalar states, $M_n^2 = \frac12 a
  n + m_\sigma^2 $. In fits I and II the uncertainties are taken as the error in the mass, or as the
  one half of the resonance width, respectively.  In both cases the lowest $n=0$ state, corresponding
  to the $f_0(600)$ resonance, is excluded from the fit.}
\end{table}

\begin{figure}[tb]
\includegraphics[width=.38\textwidth]{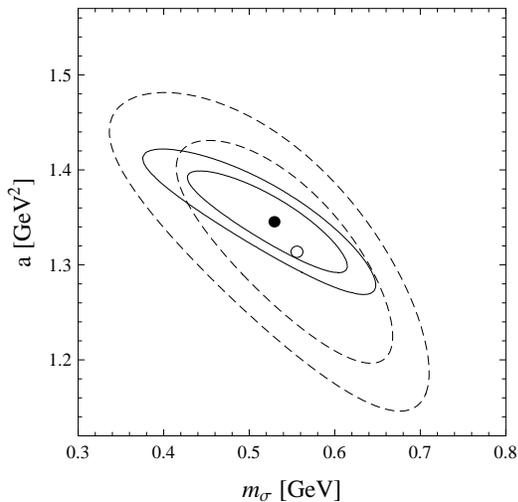}
\caption{The $\Delta \chi^2 = 2.3$ and $4.6$ contours, corresponding to the $68\%$
  and $90\%$ confidence levels in the $m_\sigma-a$ plane, for Fit~I (solid lines) and Fit~II (dashed lined). The optimum 
values are indicated with the dots, filled for Fit I, and open for Fit II.}
\label{fig:elipsefits}
\end{figure}

As is well known \cite{'tHooft:1973jz,Witten:1979kh}, at large $N_c$ mesons and glueballs are stable
states with fixed mass, $M_{S,n} \sim N_c^0$, and small decay widths
$\Gamma_{S,n} \sim 1/N_c$ and $\sim 1/N_c^2$, respectively. Moreover,
a general feature of the observed mesonic spectrum is that while the
masses grow with $n$, the widths stay constant. Therefore, one would
naively expect the large-$N_c$ approximation to work {\em better} for
the higher masses of the excited states. One might incorporate the
widths by using a particular model for the spectral density, such as
the Breit-Wigner parametrization.  Rather than using such a
parametrization for the finite width, it may be better to take the
widths themselves as an uncertainty on the value of the mass up to
${\cal O}(1/N_c)$. This is clearly an upper bound on the uncertainty
of the resonance position. Actually, the PDG resonance values are
often quoted as the Breit-Wigner values, which, depending on the
assumed background, changes from process to process and in general
leads to some model dependence. However, a rigorous description of
resonances as quantum mechanical states requires to determine them as
poles of the meson-meson scattering amplitudes in the second Riemann
sheet. Then their position in the complex plane becomes
model-independent.  For the lowest $0^{++}$ state the shift between
the Breit-Wigner and the pole values is rather large, $m_\sigma =800
{\rm ~MeV}$ vs.  $m_\sigma= 444 {\rm ~MeV}$, so the difference is
compatible with the corresponding half-width. In view of this
discussion we fit Eq.~(\ref{eq:massS}) by minimizing
\begin{eqnarray}
\chi^2 = \sum_n \left(\frac{M_{f,n}-M_S(n)}{\Gamma_{f,n}/2} \right)^2 ,
\end{eqnarray}
where we have chosen the inverse of the half-width squared as the weight.
The result of this procedure, called Fit~II, is 
\begin{eqnarray}
&& a = 1.31(12) {\rm ~GeV}^2, \; m_\sigma = 556(127) {\rm ~MeV},  \label{fitII} \\
&& \chi^2 /{\rm DOF} = 0.12 \hspace{4cm} ({\rm Fit~II}). \nonumber 
\end{eqnarray} 
The correlation plot for the $a-m_\sigma$ parameters is presented in
Fig.~\ref{fig:elipsefits} for fits I (solid lines) and II (dashed lines). We notice the compatibility of the 
two methods. 

We remark that the large-$N_c$-motivated analysis of the NN-scattering
in the $^1S_0$-channel~\cite{CalleCordon:2008eu,Cordon:2009pj} yields a quite
similar value for the lowest scalar mass, $m_\sigma = 500 (20)~{\rm MeV}$.

Although our fitting strategy was rather crude, the resulting
Regge-like spectrum of the scalar-isoscalar states turns out to be
very reasonable, given the simplicity of the mass formula of
Eq.~(\ref{eq:massS}). The crucial aspect of the foregoing analysis is
that the $f_0(600)$ resonance corresponds to the lightest state in the
radially-excited spectrum of the Regge-like scalar-isoscalar states. 
This is an important point, since scalar
glueballs~\cite{Colangelo:2007pt,Forkel:2007ru} and scalar-isoscalar
mesons~\cite{Colangelo:2008us,Afonin:2008zz} have been studied within
the AdS/CFT framework, however neglecting the possible role of the
lightest scalar in any of these approaches. The soft-wall version has
been reviewed in Ref.~\cite{Zuo:2008re,Gherghetta:2009ac}.

In Sect.~\ref{sec:grav} we will provide a mounting evidence on the
coupling of the lightest state to the gravitational form factor of
both the pion and the nucleon. This is relevant, as it shows that the
$\sigma$ state should be explicitly considered in the evaluation of
the correlation functions, where it plays an essential dynamical
role.

\section{Formalism for the effective scalar fields \label{sec:effective}}

In this Section we review the basic formalism used in the following
parts of the paper. A particular attention is paid to the
$N_c$-scaling of coupling constants of the meson and glueball
scalar-isoscalar states.

\subsection{Interpolating fields}

The first non-trivial question is to look for the interpolating QCD
operator which triggers the $J^{PC}=0^{++}$ isoscalar states from the
vacuum. On the lattice this has been done in a variety of ways, mainly
involving quark operators (such as $\bar q q$, see
also~\cite{Afonin:2008zza}). A clear candidate, often used for the
gluonium, is given by the trace of the energy-momentum tensor,
$\Theta^{\mu \nu}$, which satisfies the trace anomaly
equation~\cite{Collins:1976yq},
\begin{eqnarray} 
 \partial^\mu D_\mu &=& \Theta^\mu_\mu \equiv  \Theta  \label{def:theta} \\ 
&=& \frac{\beta(\alpha)}{2\alpha}
G^{\mu\nu a} G_{\mu\nu}^a + \sum_q m_q \left[ 1 + \gamma_m (\alpha)
\right] \bar q q . \nonumber 
\end{eqnarray} 
Here $\beta(\alpha) = \mu^2 d \alpha / d\mu^2 $ denotes the beta
function, $\alpha$ is the running coupling constant, $\gamma_m(\alpha)
= d \log m / d \log \mu^2 $ is the anomalous dimension of the current
quark mass $m$, the symbol $D_\mu$ denotes the dilatation current, and
$G_{\mu\nu}^a$ is the field strength tensor of the gluon field. The
operator $\Theta$, besides having the $J^{PC}=0^{++}$ quantum numbers,
is renorm-invariant. In addition, in the chiral limit of massless
quarks, $m_q \to 0$, it is $SU_R(2) \otimes SU_L (2)$ chirally
invariant.  This is different from the usual $\bar q q $ interpolating
field, which is not chirally invariant and mixes under the chiral
transformations with the $\bar q i \gamma_5 \vec \tau q$
operator. Actually, under the operation $q \to \gamma_5 q $, the
operator $\Theta$ is chirally even while $\bar q q$ is chirally odd.

In perturbation theory one has
\begin{eqnarray} 
\beta(\alpha) &=& - \alpha \left[\beta_0 \left( \frac{\alpha}{4\pi} \right)
+ \beta_1 \left( \frac{\alpha}{4\pi} \right)^2 + \dots \right], \nonumber \\
\gamma_m (\alpha) &=& \frac{\alpha}{4 \pi} + \dots , 
\end{eqnarray} 
where 
\begin{eqnarray} 
\beta_0 &=&\frac{11}{3}N_c-\frac{2}{3}N_f, \nonumber \\ 
\beta_1 &=& \frac{34}{3} N_c^2- \frac{13}{3}N_f N_c + \frac{N_f}{N_c},  
\end{eqnarray} 
and $N_f$ denotes the number of active flavors.
We recall that in the large-$N_c$ limit $\alpha \sim 1/N_c$ and $G^2
\sim (N_c^2 -1)$, {\em i.e.}, is proportional to the number of
gluons. Hence, in Eq.~(\ref{def:theta}) there is a contribution
scaling as $N_c^2$ and flavor-independent, as well as another
contribution, subleading in $N_c$ and scaling as $N_c N_f$ (the higher
orders in, for instance, the $\overline{MS}$ scheme, generate the
$N_f^2$ terms which are dependent on the renormalization scheme).  In
the chiral limit we only have the gluonic operator contribution to
Eq.~(\ref{def:theta}), however, still some information on the quark
degrees of freedom remains via the $\beta$-function. Thus, in the
chiral limit we may distinguish between the gluonic and quark
contributions by the large-$N_c$ and $N_f$ scaling behavior,
\begin{eqnarray}
\Theta_g \sim N_c^2 \qquad \Theta_q \sim N_c N_f.
\end{eqnarray}

\subsection{Two-point correlations \label{subsec:two-point}}

For a scalar-isoscalar particle $| n \rangle$ we have a non-vanishing matrix element to
the vacuum for the dilatation current,
\begin{eqnarray}
 \langle 0 | D^\mu | n \rangle = i q^\mu f_n ,
\end{eqnarray} 
and hence 
\begin{eqnarray}
 \langle 0 | \partial^\mu D_\mu | n \rangle = \langle 0 | \Theta | n
\rangle = m_n^2 f_n \label{dM}
\end{eqnarray}
for the on-shell particles. From now on the mass of the scalar-isoscalar states is denoted as $m_n$. 
The two-point correlation function of the energy-momentum tensor reads
\begin{eqnarray}
 \Pi_{\Theta \Theta}^{\mu \nu ; \alpha \beta } (q) &=& i \int d^4 x
 e^{i q \cdot x} \langle 0 | T \left\{ \Theta^{\mu \nu} (x)
 \Theta^{\alpha \beta} (0) \right\} | 0 \rangle . \nonumber \\ \label{PiTT}
\end{eqnarray} 
The energy momentum tensor is conserved, therefore 
\begin{eqnarray}
 \langle 0 | \Theta^{\mu \nu} | n
\rangle =  \frac13 f_n ( g^{\mu \nu} q^2 - q^\mu q^\nu ) ,
\end{eqnarray} 
where the factor of $1/3$ complies with Eq.~(\ref{dM}). After inserting a complete set of states we find  
\begin{eqnarray}
 \Pi_{\Theta \Theta}^{\mu \nu ; \alpha \beta } (q) &=& \frac19 
( g^{\mu \nu} q^2 - q^\mu q^\nu ) ( g^{\alpha \beta} q^2 - q^\alpha q^\beta ) 
 \Pi_{\Theta \Theta} (q), \nonumber \\ 
\end{eqnarray} 
with
\begin{eqnarray}
 \Pi_{\Theta \Theta} (q) &=& i \int d^4 x e^{i q \cdot x} \langle 0 |
 T \left\{ \Theta (x) \Theta (0) \right\} | 0 \rangle
 \nonumber \\ &=& \sum_n \frac{f_n^2 q^4}{m_n^2 -q^2} + {\rm c.t.}, \label{thth}
\end{eqnarray} 
where c.t. stands for the counterterms. According to the $N_c$ counting rules, the leading part of $\Pi_{\Theta \Theta}$ 
scales  as $N_c^2$, and the next-to-leading part as $N_c N_f$. 

The sum in Eq.~(\ref{thth}) runs over {\em all} $J^{PC}=0^{++}$ isoscalar states, {\em i.e.} both mesons ($m$) and glueballs ($g$). 
The difference is, however, in the scaling of the coupling constants $f_n$ with $N_c$.
If all states are contributing and no cancellations occur, 
then the glueball must have  a larger coupling, $f_{n}^2 \sim N_c^2$, whereas the
$\bar q q$-meson has $f_{n}^2 \sim N_c N_f$:
\begin{eqnarray}
& f_n \sim N_c \;\;\;\;\; &({\rm glueball}), \nonumber \\
& f_n \sim \sqrt{N_c} \;\;\; &({\rm meson}). \label{ncgm}
\end{eqnarray}  

The Operator Product Expansion (OPE) in QCD yields~\cite{
Novikov:1979va,Novikov:1981xj,Pascual:1982bv,Dominguez:1986td,Narison:1996fm}
\begin{eqnarray}
\Pi(q^2 ) = q^4 \left[ C_0 \log q^2 + \sum_{n} \frac{C_{2n}}{q^{2n}} \right], \label{qcdsr}
\end{eqnarray}   
where 
\begin{eqnarray}
C_0 &=& - \frac1{2\pi^2} (N_c^2-1) \left(\frac{\beta(\alpha)}{\alpha}\right)^2 . \label{eq:C0}
\end{eqnarray}
In the conventional treatment insisting on the presence of the {\em local} gauge-invariant operators only, 
the sum over $n$ starts with $n=2$, while the admission of the dimension-2 
condensate includes the $n=1$ term as well. In Sect.~\ref{sec:dim2} we will analyze this issue
further.

We now match the QCD result (\ref{qcdsr}) to the large-$N_c$ formula (\ref{thth}). Firstly, 
we notice that to generate the $\log q^2$ term we need
infinitely many states to contribute. Secondly, approximating the series with an integral, we find via the Euler-Maclaurin formula that
\begin{eqnarray}
f_n^2 / (d  m_n^2 /dn ) \to {\rm const.}
\label{eq:f-asymp}
\end{eqnarray} 
For the Regge spectrum in Eq.~(\ref{eq:massS}) this specifically means
that $f_n^2$ does not asymptotically depend on $n$. 

The asymptotic $q^4 \log q^2$ behavior allows one to write down a 
twice-subtracted dispersion relation,
\begin{eqnarray}
\Pi(q^2 ) = \Pi(0)+ \Pi'(0) q^2 + \frac{q^4}{\pi} \int  \frac{dt}{t^2}
\frac{{\rm Im} \Pi(t)}{t-q^2 - i\epsilon}.
\end{eqnarray}   
In our case, the spectral density is 
\begin{eqnarray}
\frac1{\pi} {\rm Im} \Pi (s) = \sum_n f_n^2 m_n^4 \delta( m_n^2 -s ),
\end{eqnarray}   
hence we can write 
\begin{eqnarray}
 \Pi_{\Theta \Theta} (q)  = \sum_n \frac{f_n^2 m_n^4}{m_n^2 -q^2} .
\end{eqnarray} 
We have imposed the condition to have a well
behaved object at high energies. With this prescription
the low-energy theorem~\cite{Novikov:1981xj,Donoghue:1991qv,Narison:1988ts},
is written as 
\begin{eqnarray}
\Pi_{\Theta \Theta} (0) = \sum_n f_n^2 m_n^2 = - 4 \langle \Theta \rangle = -16 B_v, \label{bv}
\end{eqnarray} 
where $B_v$ is the energy density of the vacuum.  The first derivative with respect to $q^2$ yields  
\begin{eqnarray}
\Pi_{\Theta \Theta}' (0) = \sum_n f_n^2 \sim \int d^4 x x^2 \langle
\Theta (x) \Theta (0) \rangle,
\label{eq:sumfn^2}
\end{eqnarray} 
a dimension-2 object which cannot be expressed as an expectation
value of a local operator. The second derivative is equal to 
\begin{eqnarray}
\frac12 \Pi_{\Theta \Theta}'' (0) = \sum_n \frac{f_n^2}{m_n^2}, \label{eq:2sumfn^2}
\end{eqnarray} 
a dimensionless object which corresponds to the graviton wave-function
renormalization and which may be divergent. In general, if we use the
fact that $f_n$ becomes asymptotically $n$-independent and $m_n^2 \sim
n$, as suggested by the radial Regge trajectories, the sums (\ref{bv})
and (\ref{eq:sumfn^2}) need regularization (see Sect.~\ref{sec:dim2}
for an application of the $\zeta$-function) and the naively positive
combination of Eq.~(\ref{eq:sumfn^2}) may in fact turn into a
vanishing or even negative contribution.

\subsection{Effective Lagrangeans}

The results of the previous subsections can be directly translated
into the language of the effective Lagrangeans.  In particular, we may
write the PCDC (Partial Conservation of the Dilaton Current) equation,
\begin{eqnarray}
D^\mu (x) = \sum_n f_n \partial^\mu \varphi_n(x) + {\cal O}(\varphi^2),
\end{eqnarray} 
and hence for the on-shell particles we may use the equations of motion,
$(-\partial^\mu \partial_\mu - m_n^2 ) \varphi_n =0$, to obtain
\begin{eqnarray}
\Theta (x) = -\sum_n f_n m_n^2 \varphi_n(x) + {\cal O}(\varphi^2).
\end{eqnarray} 
For the off-shell particles we need to introduce the improved energy
momentum tensor~\cite{Callan:1970ze}, which has a better high energy
behavior, but the final result is the same.

The above properties may be described with a suitable extension of
the effective dilatonic
Lagrangeans~\cite{Schechter:1980ak,Migdal:1982jp,Gomm:1985ut,Ellis:1984jv}
of the form
\begin{eqnarray}
{\cal L} (x) = \sum_n \left[ \frac12 \partial^\mu \Phi_n\partial_\mu
\Phi_n - V(\Phi_n) \right].
\end{eqnarray} 
Here, the potential 
\begin{eqnarray}
V(\Phi_n) = 
\frac{m_n^2}{8 f_n^2} \Phi_n^4 \left[ \log\left( \frac{\Phi_n}{f_n}
\right)^2 - \frac12 \right]
\end{eqnarray} 
fulfills the conditions that the minimum is at $\langle \Phi_n \rangle
=f_n$ and the mass is $V''(f_n) = m_n^2$. The physical field
$\varphi_n$ is defined by $\Phi_n = f_n + \varphi_n $.  The value at
the minimum is
\begin{eqnarray}
\epsilon_v = \sum_n V(f_n) =  -\frac1{16} \sum_n m_n^2 f_n^2, \label{epsilon}
\end{eqnarray} 
which corresponds to the vacuum energy density (\ref{bv}).

The coupling to the pion can be considered by working in the non-linear representation 
and with the interaction Lagrangean
\begin{eqnarray}
{\cal L}_{\pi} &=& \frac{f_\pi^2}{4} \sum_n   \frac{g_{n\pi \pi}}{f_n} \Phi_n^2    \langle
\partial^\mu U^\dagger \partial_\mu U \rangle , \label{lagpi}
\end{eqnarray}
where $U(x)= e^{i \vec \tau \cdot \pi / f_\pi}$ is the non-linearly
transforming pion field and $\langle . \rangle $ represents the trace
in the isospin space. In the vacuum, $\Phi_n=f_n$, we get
\begin{eqnarray}
{\cal L}_{\pi}^{(0)} &=& \frac{f_\pi^2}{4} \sum_n {g_{n\pi \pi}}{f_n}    \langle
\partial^\mu U^\dagger \partial_\mu U^\dagger \rangle , \label{lagpi0}
\end{eqnarray}
which leads to the condition
\begin{eqnarray}
\sum_n {g_{n\pi \pi}}{f_n} =1. \label{eq:cond1}
\end{eqnarray}
Since ${\cal L}_{\pi} \sim N_c$ and $f_\pi \sim \sqrt{N_c}$, it follows (under the assumption that all states are 
contributing in the large-$N_c$ limit and no cancellations occur) 
that $g_{n\pi \pi} f_n \sim N_c^0$, or, due to Eq.~(\ref{ncgm}),
\begin{eqnarray}
& g_{n \pi \pi} \sim 1/N_c \;\;\;\;\; &({\rm glueball}), \nonumber \\
& g_{n \pi \pi} \sim 1/\sqrt{N_c} \;\;\; &({\rm meson}). \label{ncggm}
\end{eqnarray}  
A construction of the chiral Lagrangean in the linear realization is
presented in Appendix~\ref{sec:lin-sig}.

The scalar fields $\Phi_n$ are invariant under chiral
transformations.\footnote{This is unlike the standard assignment of
  the linear sigma-model where one takes $(\sigma, \vec \pi)$ as
  chiral partners in the $(1/2,1/2) $ representation of the chiral
  $SU(2)_R \otimes SU(2)_L$ group. See also
  Appendix~\ref{sec:lin-sig}.}
Lagrangean (\ref{lagpi}) is nothing else but a generalization of the
dilaton Lagrangean of Refs.~\cite{Schechter:1980ak,Gomm:1985ut} for
the case of many dilatons. We treat here {\em all} scalars on equal
footing, as suggested by the fact that they all fit into the same
radial Regge trajectory.

The tree-level decay width of the scalar $n$ into two pions is given
by the expression
\begin{eqnarray}
&&\Gamma ( \Phi_n \to \pi \pi ) \equiv \Gamma_{n \pi \pi} =\label{decay:gamma} \\
&&\frac{3 g_{n \pi \pi}^2 m_n^2}{32\pi} \left[1- \frac{2
m_\pi^2}{m_n^2}\right]^2 \sqrt{m_n^2 - 4 m_\pi^2} 
\sim  \frac{3 g_{n \pi \pi}^2 m_n^3}{32\pi}. \nonumber
\end{eqnarray} 
Because of the scaling (\ref{ncggm})
\begin{eqnarray}
& \Gamma_{n \pi \pi} \sim 1/N_c^2 \;\;\;\;\; &({\rm glueball}), \nonumber \\
& \Gamma_{n \pi \pi} \sim 1/N_c \;\;\; &({\rm meson}), \label{ncGgm}
\end{eqnarray}  
which displays the weaker coupling of the glueballs to the pion compared to the mesons.

The  Lagrangean describing the interaction 
of the nucleon with the scalar-isoscalar particles is 
\begin{eqnarray}
{\cal L}_N = \bar N \left( i \slashchar{\partial} - \sum_n g_{nNN}
\Phi_n - \frac{g_A}{f_\pi} u^\dagger \gamma_5 \slashchar{\partial} u \right) N, \label{phiN}
\end{eqnarray}
with $u^2 = U^5=e^{i \gamma_5 \vec \tau \cdot \pi / f_\pi}$. The nucleon field $N$ transforms non-linearly, but we can return to
the linear realization by undoing the chiral transformation, $\Psi = u N$,
which effectively replaces the nucleon mass $M_N$ by the
combination $\sum_n g_{nNN} \Phi_n$. This yields the 
generalized Goldberger-Treiman relation for the scalar-isoscalar channel,
\begin{eqnarray}
M_N = \sum_n   g_{n NN} f_n, \label{gtc}
\end{eqnarray} 
derived by Carruthers in the early seventies~\cite{Carruthers:1971vz}. 
Again, under the assumption that all states contribute to (\ref{gtc}) and no cancellations occur, we find
\begin{eqnarray}
& g_{n NN} \sim N_c^0 \;\;\;\;\; &({\rm glueball}), \nonumber \\
& g_{n NN} \sim \sqrt{N_c} \;\;\; &({\rm meson}). \label{ncgNN}
\end{eqnarray}  

Lagrangean (\ref{phiN}) fulfills also the standard Goldberger-Treiman
relation, $g_A M_N = f_\pi g_{\pi NN}$. In addition, it is
scale-invariant, such that as a consequence that $D^\mu$ is the
Noether current generating dilatations we have the
identity~\cite{Schechter:1980ak,Migdal:1982jp,Gomm:1985ut,Ellis:1984jv}
\begin{eqnarray}
\partial^\mu D_\mu &=& \Theta = \sum_n \left[ 4 V(\Phi_n) - \Phi_n
V'(\Phi_n) \right] \nonumber \\
&=& - \sum_n \frac{m_n^2}{4 f_n^2} \Phi_n^4. 
\end{eqnarray}
Upon subtraction of the vacuum part and using that $\Phi_n = f_n +
\varphi_n$ it follows that
\begin{eqnarray}
\langle N | \Theta(0) | N \rangle &=& -\sum_n m_n^2 f_n \langle N |
\varphi_n | N \rangle \nonumber \\ &=& \sum_n f_n g_{n NN}, ~~
\end{eqnarray}
in agreement with (\ref{gtc}). In other words, the pole mass and the energy-momentum mass are the same, as required by consistency. 

The contribution of the scalar-isoscalar degrees of freedom  to the $\pi N$ scattering is given by 
the formula
\begin{eqnarray}
T_{\pi N}^{ab} = \delta^{ab} \, \bar u (p') u(p) \, \sum_n \frac{q^2 g_{n\pi\pi} g_{n NN}}{m_n^2-q^2},
\end{eqnarray} 
which vanishes at the threshold $q=0$, in agreement with the
Weinberg-Tomozawa result. According to Eqs.~(\ref{ncggm}) and
(\ref{ncGgm}), the meson contribution scales as $N_c^0$ and the
glueball contribution as $1/N_c$, hence is relatively suppressed.

Finally, the nucleon-nucleon central potential stemming from the
exchange of the $J^{PC}=0^{++}$ isoscalar particles reads in the
large-$N_c$ limit
\begin{eqnarray}
V_C (|{\bf q}|) = \sum_n \frac{g_{nNN}^2}{{\bf q}^2 +  m_n^2}.  
\end{eqnarray} 
Passing to the coordinate space we obtain the standard Yukawa potential,
\begin{eqnarray}
V_C (r) = - \sum_n \frac{g_{nNN}^2}{4 \pi} \frac{e^{-m_n r}}{r} 
\end{eqnarray} 
The large-$N_c$ counting rules (\ref{ncGgm}) yield $V_C \sim N_c$ from the meson exchange, 
while the glueball exchange produces a subleading contribution $\sim N_c^0$.
Other tensorial components of the NN potential scale as
$\sim N_c$ or $\sim 1/N_c$~\cite{Kaplan:1996rk,Banerjee:2001js}.  

The $N_c$ scaling of various quantities is collected for convenience
in Table~\ref{tab:nc}.
\begin{table}[tb]
\caption{$N_c$-scaling of various quantities. \label{tab:nc}}
\begin{center}
\begin{tabular}{|c|c|c|} 
\hline
quantity & glueball & $q \bar q$ meson \\ \hline
$m_n$ & 1  & 1 \\
$f_n$    & $N_c$ & $\sqrt{N_c}$ \\
$\Gamma_{n\pi\pi}$ & $1/{N_c^2}$  & $1/N_c$ \\
$g_{n\pi\pi}$ & $1/N_c$  & $1/\sqrt{N_c}$ \\
$g_{nNN}$ & $1$  & $\sqrt{N_c}$ \\
\hline
\end{tabular}
\end{center}
\end{table}

\section{Gravitational form factors \label{sec:grav}}

In this section we use the lattice data for the gravitational form
factors of the pion and nucleon to place further bounds on the mass of
the $\sigma$ meson. Our analysis is based on the fact that the spin-0
channel is dominated by the scalar-isoscalar mesons.

\subsection{Pion}

The gravitational form factors of the pion are defined via the matrix element of the energy-momentum tensor between 
the pion states, 
\begin{eqnarray}
&& \!\!\!\!\!\! \langle \pi^b(p') \mid \Theta^{\mu \nu}(0) \mid \pi^a(p) \rangle = \nonumber\\ && 
\!\!\!\!\!\!\! \frac{1}{2}{\delta^{ab}}\left [ (g^{\mu \nu}q^2- q^\mu q^\nu)
\Theta_1(q^2)+ 4 P^\mu P^\nu \Theta_2(q^2) \right ], 
\label{eq:pion-em}
\end{eqnarray}
where $P=\frac{1}{2}(p'+p)$, $q=p'-p$, and $a, b$ are the isospin
indices. The trace part of this form factor corresponds to the coupling of the spin-0, while
the traceless part to spin-2 states. In the present analysis we are interested in the former.
The spin-0 gravitational form factor is given by
\begin{eqnarray}
\langle \pi^b (p') | \Theta (0) | \pi^a (p) \rangle = \delta^{ab} \Theta_{\pi} (q^2),
\end{eqnarray} 
where 
\begin{eqnarray}
\Theta_\pi (q^2) = \frac{3}{2} q^2 \Theta_1 (q^2) + \frac12 ( 4m_\pi^2 -q^2) \Theta_2 (q^2). 
\label{eq:pion-spin-0}
\end{eqnarray} 
Due to chiral symmetry constraints, this form factor
satisfies~\cite{Novikov:1981xj,Donoghue:1991qv,Narison:1988ts},
\begin{eqnarray}
\Theta_{\pi} (q^2) = q^2 + {\cal O} (p^4).  
\end{eqnarray} 
Two low energy theorems follow:
\begin{eqnarray}
\Theta_{\pi} (0) &=& 0 , \\
\Theta_{\pi}' (0) &=& 1.  
\end{eqnarray} 
These conditions can be achieved by assuming a derivative coupling
\begin{eqnarray}
\langle \pi^a (p') | \Phi_n | \pi^b(p) \rangle = \delta^{ab}
\frac{g_{n \pi \pi} q^2}{m_n^2-q^2},
\end{eqnarray} 
written as 
\begin{eqnarray}
\Theta_{\pi} (q^2) = \sum_n \frac{g_{n \pi \pi} f_n q^2
m_n^2}{m_n^2-q^2},
\end{eqnarray} 
and thus
\begin{eqnarray}
\sum_n f_n g_{n \pi \pi} = 1, \label{rel1}
\end{eqnarray} 
exactly as Eq.~(\ref{eq:cond1}).  Since according to
Table~\ref{tab:nc} we have the scaling $f_n g_{n \pi \pi} \sim N_c^0$,
both mesons and glueballs may contribute to relation (\ref{rel1}) on
equal footing.

\begin{figure}[tb]
\begin{center}
\includegraphics[width=.41\textwidth]{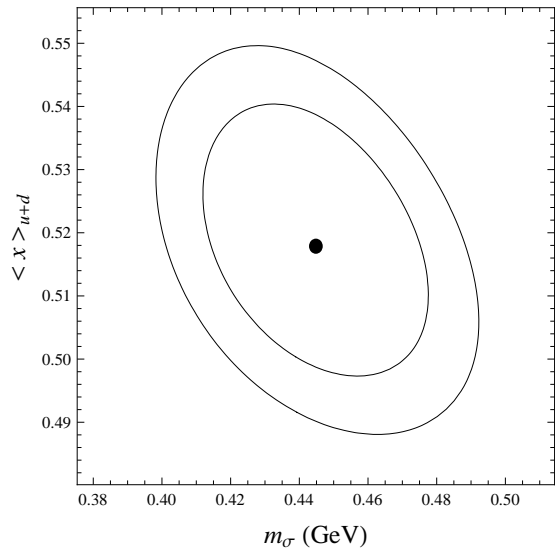} 
\end{center}
\caption{The $\Delta \chi^2 = 2.3$ and $4.6$ contours, corresponding
  to the $68\%$ and $90\%$ confidence levels, in the $m_\sigma-\langle
  x \rangle_{u+d}$ plane. The fit is obtained from the lattice data
  for the pion spin-0 gravitational form factor, as described in the
  text.}
\label{fig:piel}
\end{figure}

At large momenta the form factor behaves as 
\begin{eqnarray}
\Theta_{\pi} (q^2) \to -\sum_n g_{n \pi \pi} m_n^2 f_n - \frac{\sum_n g_{n
\pi \pi} m_n^4 f_n}{q^2} + \dots
\end{eqnarray} 
If we require the constant term to vanish, we get 
\begin{eqnarray}
\sum_n f_n g_{n \pi \pi} m_n^2 = 0.  \label{sr:0}
\end{eqnarray} 
Clearly, cancellations are necessary and we need at least two contributing states to satisfy the sum rule~(\ref{sr:0}). This 
high-energy relation plays a similar role in the scale-invariance considerations
to the second Weinberg sum rule in the chiral framework. 

\begin{figure}
\includegraphics[width=.47\textwidth]{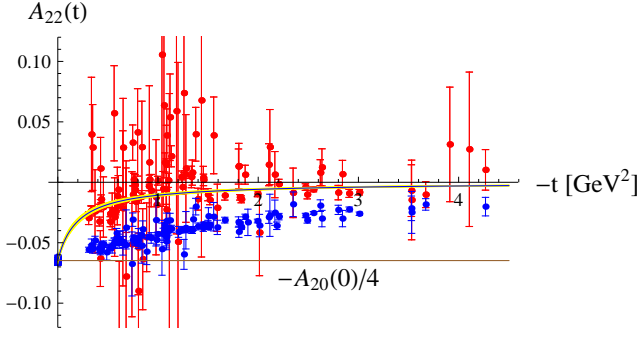} 
\caption{(Color online) 
Spin-0 gravitational form factor of the pion, $A_{22} (t)$, from
the lattice calculation of Refs.~\cite{Brommel:2007xd,Brommel:PhD}
extrapolated to the physical pion mass (the points with the large error bars), together with the monopole
fit with parameters (\ref{opti2}), indicated by the line. The band indicates the uncertainty in the model parameters. The low energy theorem
fixes the value of the $A_{22}$ form factor at $t=0$. The lower data points show the data for $-A_{20}/4$.  
\label{fig:A22pion}}
\end{figure}

The only determination of the pion gravitational
form factor to date is the lattice calculation of Brommel {\it et al.}~\cite{Brommel:2007xd,Brommel:PhD}. 
The data go up to $t=-4 {\rm ~GeV}^2$, however with substantial errors. The quark contribution to
the energy momentum tensor, considered in~\cite{Brommel:2007xd,Brommel:PhD},  is
\begin{eqnarray}
\Theta^{\mu \nu}_q (x)= \frac14 \bar q(x) \left[\gamma^\mu \, i \lrd\/^\nu + \gamma^\nu\, i \lrd\/^\mu \right] q (x),
\end{eqnarray} 
from which a decomposition similar to Eq.~(\ref{eq:pion-em}) and the
corresponding spin-0 component Eq.~(\ref{eq:pion-spin-0}) follow.  The
data points are shown in Fig.~\ref{fig:A22pion}. Following the
standard notation for the moments of the pion GPD's \cite{Broniowski:2009zh} we introduce
\begin{eqnarray}
&& A_{20}(t)=\frac{1}{2}\Theta_1^q(t), \;\; A_{22}(t)=-\frac{1}{2}\Theta_2^q(t), \label{lowff}
\end{eqnarray}
where the symbols $\Theta_i^q$ denote the quark parts of the gravitational form
factors of Eq.~(\ref{eq:pion-em}). A monopole fit has been undertaken,
yielding $m_\sigma = 0.89(27)(9) {\rm ~MeV}$
and $A_{22}(0)=-0.076(5)$ ({\em cf.} Table~C.8 on p.~105 of
Ref.~\cite{Brommel:PhD}).  By using several parameterizations it has
been noted that the precise form of the function cannot be pinned down from
the data unambiguously due to large uncertainties in $A_{22}(t)$.  In
the following we will use the low-energy theorem
\begin{eqnarray}
A_{22}(0)=-\frac{1}{4}A_{20}(0), \label{letA2}
\end{eqnarray}
which provides a relatively accurate fixing of $A_{22}$ at the origin and greatly helps the 
regression analysis.

In terms of the form factors related to the Generalized Parton Distributions (GPD) of
Ref.~\cite{Brommel:2007xd,Brommel:PhD}, the quark contribution to the
gravitational form factor reads
\begin{eqnarray}
\Theta_\pi^q ( t) = - 4 t A_{22}^q (t) 
\end{eqnarray} 
Due to the multiplicative QCD evolution one has 
\begin{eqnarray}
\Theta_\pi^q ( t, \mu) = \langle x \rangle_q^\pi (\mu) \Theta_\pi (t),
\end{eqnarray} 
where the (valence) quark momentum fraction depends on the
renormalization scale $\mu$. Its leading-order perturbative evolution reads
\begin{eqnarray}
R=\frac{ \langle x \, \rangle_q (\mu) } { \langle x \, \rangle_q (\mu_0)  } = \left( \frac{\alpha(\mu)}
{\alpha(\mu_0) } \right)^{\gamma_1^{\rm (0)} / (2 \beta_0) },
\end{eqnarray} 
where the anomalous dimension is  $ \gamma_1^{\rm (0)} / (2\beta_0) = 32/81 $ for $N_F=N_c=3$. The QCD 
running coupling constant is equal to
\begin{eqnarray}
\alpha(\mu)=\frac{4\pi}{\beta_0 \log(\mu^2/\Lambda^2_{\rm QCD})}, \label{gambe}
\end{eqnarray}
where we take $\Lambda_{\rm QCD} = 226~{\rm MeV}$ for $N_c=N_f=3$. 

Phenomenologically, it is known from the Durham
group analysis~\cite{Sutton:1991ay}, based mainly on the E615
Drell-Yan data~\cite{Conway:1989fs} and the model assumption that the sea
quarks carry $10-20\%$ fraction of the momentum, that $\langle x
\rangle_{u+d}^\pi = 0.47 (2)$ at the scale $\mu= 2~{\rm GeV}$. The
analysis of the Dortmund group~\cite{Gluck:1999xe}, based on the
assumption that the momentum fraction carried by the valence quarks in
the pion coincides with that in the nucleon, yields $\langle x \rangle
_{u+d}^\pi = 0.4$ at $\mu= 2~{\rm GeV}$.  The lattice data at the
lattice spacing $a_{\rm lat}= 0.1 {\rm fm}$ as well as a recent chiral
quark model calculation~\cite{Broniowski:2008hx} support this view.

We recall that the large-$N_c$ analysis implies sums of monopoles in the form factors. 
The largest available momentum transfer, $t=- 4 {\rm ~ GeV}^2$, obtained in Ref.~\cite{Brommel:PhD},
suggests that some information on the contribution of the excited states might be
extracted. Therefore, following the approach already used for the
electromagnetic pion form factor in Ref.~\cite{Dominguez:2001zu} (see
also \cite{RuizArriola:2008sq}), we have attempted a Regge-like
fit, 
\begin{eqnarray}
\Theta_\pi (t)= t f_b(t) ,
\end{eqnarray} 
including infinitely many states, of the form
\begin{eqnarray}
f_b(t)=\frac{B ( b -1 , \frac{M^2 - t}{a/2} )}{B ( b -1 , \frac{M^2}{a/2} )}, \label{f}
\end{eqnarray} 
with $B(x,y)= \Gamma(x) \Gamma(y) /\Gamma(x+y)$ denoting the Euler Beta function.
The function (\ref{f}) fulfills the normalization condition
\begin{eqnarray}
f_b(0)=1.
\end{eqnarray}  
For  $x \gg y $  one has  $B(x,y) \sim \Gamma(y ) x^{-y}$, hence in the 
asymptotic region of $ M^2 - t \gg (b-1) a$ we find
\begin{eqnarray}
f_b(t) \sim \frac{\Gamma\left( \frac{M^2}{a/2} +b-1 \right)}{\Gamma\left(
\frac{M^2}{a/2}\right)} \left( \frac{a/2}{M^2-t} \right)^{b-1}.
\end{eqnarray} 
The result for $a=1.31$ is 
\begin{eqnarray}
\langle x \rangle_{u+d}^\pi = 0.52(3), \; m_\sigma = 495^{+250}_{-135}
        {\rm ~MeV}, \; b=2.24^{+1.56}_{-0.55}. \nonumber \\ 
\end{eqnarray} 
As we can see, the result is fully compatible with the monopole
($b=2$) and at present the large errors in the lattice data wash out
any insight from the excited scalar spectrum, despite the large
momenta\footnote{We note that a similar fit \cite{RuizArriola:2008sq}
  to the vector form factor using $a=1.2(1) {\rm ~GeV}^2$ yields
  $m_\rho= 775(15) {\rm ~MeV}$ and $b=2.14(7)$, which can be
  distinguished from a simple monopole fit at the
  two-standard-deviation level.}.

Thus, we restrict ourselves to a simple monopole fit
\begin{eqnarray}
\Theta_\pi^q (t) = \langle x \rangle_{u+d}^\pi \frac{ t \,
  m_\sigma^2}{m_\sigma^2-t},
\end{eqnarray}
which yields $\chi^2/{\rm DOF} \sim 2.4$.  Actually the large $\chi^2$
is due to incompatible values of nearby points. In such a situation,
in order to obtain reliable estimate of the model parameters, we
rescale the errors by a factor of 1.5 to make them mutually
compatible.  Moreover, we enforce the low energy theorem,
Eq.~(\ref{letA2}) as a constraint -- a possibility not directly
considered in Ref.~\cite{Brommel:2007xd,Brommel:PhD}. As a result we
get $\chi^2 /{\rm DOF} \simeq 1$ (after the mentioned rescaling of the
data errors) and the optimum values
\begin{eqnarray}
\langle x \rangle_{u+d}^\pi = 0.52(2) \qquad m_\sigma = 445(32) {\rm ~MeV} . \label{opti2}
\end{eqnarray}
In Fig.~\ref{fig:piel} we present the corresponding correlation ellipse. The
gravitational form factor $\Theta_0(t)$ at the optimum values of the parameters (\ref{opti2}) is presented in
Fig.~\ref{fig:A22pion}. 

\subsection{Nucleon}

The scalar component of the nucleon gravitational form factor is
\begin{eqnarray}
\langle N(p') | \Theta (0) | N(p) \rangle = \bar u(p') u(p)
\Theta_{N} (q^2),
\end{eqnarray} 
where $u(p)$ and $u(p')$ are nucleon Dirac spinors. In the meson-dominance
approximation it can be written as
\begin{eqnarray}
\Theta_N (q^2) = \sum_n  g_{n NN}  \frac{ f_n  m_n^2}{m_n^2-q^2} .
\end{eqnarray} 
The nucleon mass is given by Eq.~(\ref{gtc}).

The quark contributions to the GPD's of the nucleon have been
determined on the lattice~\cite{Gockeler:2003jfa,Hagler:2007xi} for
masses as small as twice the physical pion mass. The decomposition
corresponding to the energy-momentum tensor reads
\begin{eqnarray}
\langle p^\prime| \Theta_{\mu\nu}^{q} |p\rangle &=& \bar u(p^\prime)\biggl[
    A^{q}_{20}(t)\,\frac{\gamma_\mu P_\nu+\gamma_\nu P_\mu}{2} \nonumber \\ 
&+&  B^{q}_{20}(t)\,\frac{i(P_{\mu}\sigma_{\nu\rho}+P_{\nu}\sigma_{\mu\rho})
    \Delta^\rho}{4M_N}  \nonumber \\
    &+& C^{q}_{20}(t)\,\frac{\Delta_\mu\Delta_\nu-g_{\mu\nu}\Delta^2}{M_N}
    \biggr]u(p),  \label{emt:N} 
\end{eqnarray}
where $\sigma^{\mu\nu}= \frac{i}{2} [ \gamma^\mu, \gamma^\nu] $
(the Bjorken-Drell notation), the scalar functions are moments of the GPD's,
the momentum transfer is denoted as $\Delta=p'-p$, and 
the average nucleon momentum is $P= (p'+p)/2$. Taking the trace and applying the
Gordon identity, $2 M_N \bar u(p') \gamma^\mu u(p) = \bar u(p') ( i
\sigma^{\mu\rho} \Delta_\rho + 2P^\mu ) u(p) $, as well as the Dirac equation,
$(\slashchar{p}-M_N)u(p)=0$ and $\bar u(p')(\slashchar{p'}-M_N)=0$, we obtain
the following expression for the spin-0 gravitational form factor of the nucleon:
\begin{eqnarray}
\!\!\!\Theta_N^q( t) = M_N \left[ A_{20}^q (t) + \frac{t}{4M_M^2} B_{20}^q (t) -
\frac{3t}{M_N^2} C_{20}^q (t) \right] .
\label{eq:thetaNq}
\end{eqnarray} 
Due to the multiplicative character of the QCD evolution for
the conserved energy-momentum tensor operator, one has
\begin{eqnarray}
\sum_q \Theta_N^q ( t) = \sum_q \langle x \rangle_q^N \Theta_N (t) \equiv 
\langle x \rangle_{u+d} \Theta_N(t). \label{evol:x:N}
\end{eqnarray} 

\begin{figure}
\includegraphics[width=.47\textwidth]{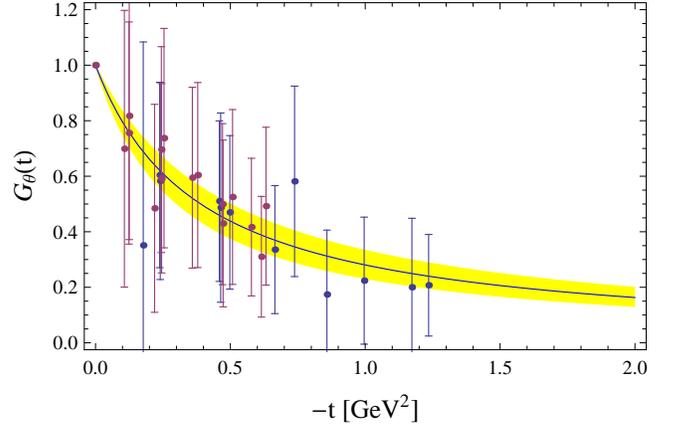} 
\caption{(Color online) 
Spin-0 gravitational form factor of the nucleon, $G_\theta
  (t)$, obtained from the lattice simulations of
  Refs.~\cite{Hagler:2007xi} at the pion masses $m_\pi = 352$ and $356
  {\rm ~MeV}$, together with the monopole fit with mass $m_\sigma =
  624 {\rm ~MeV}$ (solid line). The band indicates the uncertainty in
  $m_\sigma$ of $78$~MeV obtained with the $\chi^2$ method.
\label{fig:Thetanucleon}}
\end{figure}

\begin{figure}
\begin{center}
\includegraphics[width=.41\textwidth]{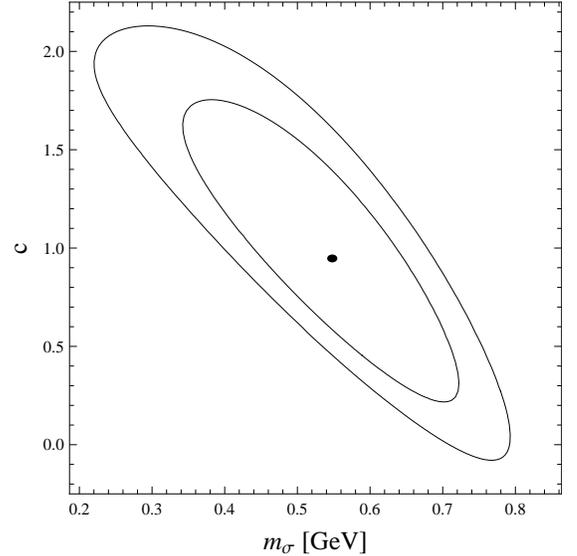} 
\end{center}
\caption{The $\Delta \chi^2 = 2.3$ and $4.6$ contours, corresponding
  to the $68\%$ and $90\%$ confidence levels in the $m_\sigma-c$
  plane. The fit is obtained from the nucleon spin-0 gravitational
  form factor obtained from the LHPC data~\cite{Hagler:2007xi}. See
  the text for details.
\label{fig:ms-c}}
\end{figure}

The combination (\ref{eq:thetaNq}) can be directly constructed from
the results of the LHPC Collaboration~\cite{Hagler:2007xi}, where the data
points for the $A_{20}$, $B_{20}$, and $C_{20}$ form factors are
provided for six values the pion mass: $m_\pi=757$, $681$, $595$,
$495$, $356$, and $352$~MeV.  The values of the corresponding nucleon masses are
$M_N=1565$, $1489$, $1379$, $1292$, $1216$, and $1158$~MeV.  We add the
statistical errors in quadrature. The result for the two lowest pion
masses is shown in Fig.~\ref{fig:Thetanucleon}. We note that while the
$A_{20}(t)$ function is determined to a better than $10\%$ accuracy, the
large errors in $\Theta_N^q$ originate from the less accurate $B_{20}(t)$ and
$C_{20}(t)$ form factors.

Similarly to the pion case we have tried to deduce some information 
from the high-energy region, $t \sim -1.2 {\rm ~GeV}^2$, by attempting the Regge form 
\begin{eqnarray}
\Theta_N (t)= M_N \frac{B ( b -1 , \frac{M^2 - t}{a/2} )}{B ( b -1 , \frac{M^2}{a/2} )}. \label{fff}
\end{eqnarray} 
Unfortunately, again the large errors in the data do not allow to fix the $b$ parameter 
unambiguously, and it is compatible with $b=2$. Thus we stick to the simple monopole form, 
\begin{eqnarray}
\Theta_N^q (t)= M_N \langle x \rangle_{u+d}^N \frac{m_\sigma^2}{m_\sigma^2-t}. 
\end{eqnarray}

Since most of the lattice data are accumulated at pion masses significantly above the physical value, extrapolation to the physical point is necessary. 
At low values of the pion mass and $t$ the chiral corrections are expected to be significant.\footnote{Experience with the 
vector form factor shows that the low momentum
region is quite sensitive to chiral corrections, whereas the
intermediate energy region is better described by the large-$N_c$ dynamics.} 
Not entering these intricacies, we assume, following many lattice studies, 
the simple dependence of $m_\sigma$ on $m_\pi$: 
\begin{eqnarray}
m_\sigma^2(m_\pi)=m_\sigma^2+c\left  ( m_\pi^2-m_{\pi,{\rm phys}}^ 2 \right ). \label{msmpi}
\end{eqnarray}
Hence $m_\sigma$ (without an argument) denotes the value at the physical point. 
Similarly, 
\begin{eqnarray}
\langle x \rangle_{u+d}^N(m_\pi) =\langle x \rangle_{u+d}^N +d \left  ( m_\pi^2-m_{\pi,{\rm phys}}^ 2 \right ). \label{xmpi}
\end{eqnarray}

We use simultaneously the form factors (\ref{eq:thetaNq}) for the six
pion masses from LHPC~\cite{Hagler:2007xi}. There are several ways one
can perform the fit, all leading to similar but not identical
results. Our procedure is as follows: We first normalize the form
factors to unity at $t=0$, by dividing a data set at a particular
$m_\pi$ by the value of the point at $t=0$.  That way we get rid of
the dependence on $\langle x \rangle_{u+d}^N$ and can carry out a
two-parameter fit, with the parameters of Eq.~(\ref{msmpi}).  The
$\chi^2$ method yields
\begin{eqnarray}
\ m_\sigma = 550^{+180}_{-200} {\rm MeV} , \;\;\; c=0.95^{+0.80}_{-0.75}, \label{best:ms-c}
\end{eqnarray}
with  $\chi^2 /{\rm DOF}= 5.2/(87-2)$. 
The resulting correlation plot is shown in Fig.~\ref{fig:ms-c}

\begin{figure}
\includegraphics[width=.47\textwidth]{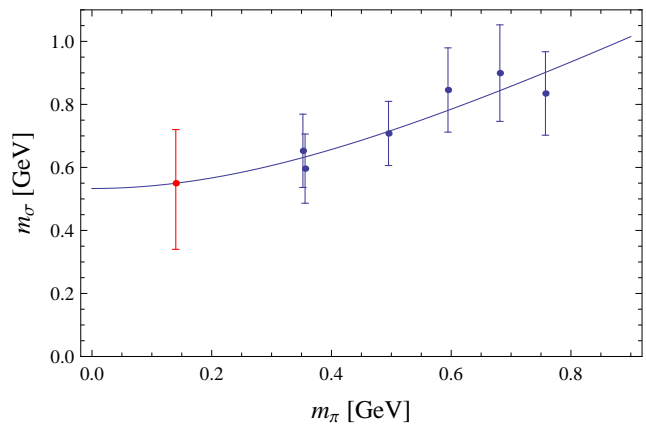} 
\caption{(Color online) 
Dependence of the $\sigma$ mass on $m_\pi$ inferred from the  LHPC data~\cite{Hagler:2007xi}. The line shows the best fit 
of Eq.~(\ref{best:ms-c}). The point at physical $m_\pi$ indicates the result of the extrapolation.
\label{fig:ms_mpi}}
\end{figure}

\begin{figure}
\includegraphics[width=.47\textwidth]{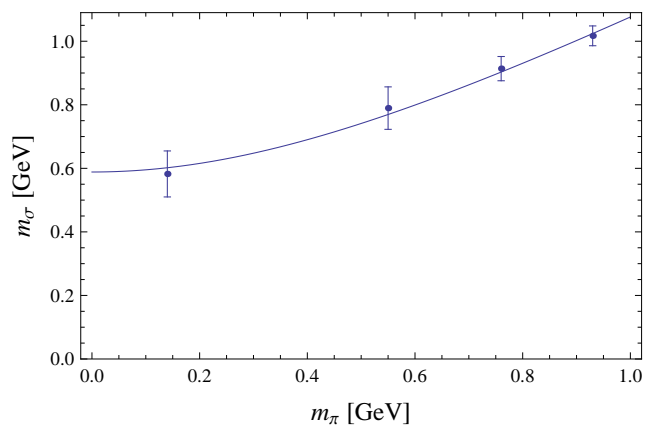} 
\caption{(Color online) Dependence of the $\sigma$ mass on $m_\pi$ inferred from the
  QCDSF data~\cite{Gockeler:2003jfa}. The line indicates the best fit
  of Eq.~(\ref{best:sf-ms-c}).
\label{fig:sf_ms_mpi}}
\end{figure}

One may perform separately the fit to the normalized form factors at
each value of $m_\pi$.  An example for the two lowest pion masses,
$356$ and $352$~MeV, is shown in Fig.~\ref{fig:Thetanucleon}. The band
indicates the uncertainty in $m_\sigma$ obtained with the $\chi^2$
method.  The result of the procedure at all values of $m_\pi$ is shown
in Fig.~\ref{fig:ms_mpi}, together with the best fit curve with
parameters (\ref{best:ms-c}).  The point at the physical value of
$m_\pi$ indicates the resulting extrapolation for the physical value of
$m_\sigma$, Eq.~(\ref{best:ms-c}).

Taking the  values of the unnormalized form factors at $t=0$ we may compute the momentum fraction carried by the quarks as a 
function of $m_\pi$. The result of the $\chi^2$ fit is
\begin{eqnarray}
\langle x \rangle_{u+d}^N = 0.447(14), \;\;\; d=-0.13(3).
\end{eqnarray}

In the case of the lattice calculations of the QCDSF
Collaboration~\cite{Gockeler:2003jfa}, five points have been provided
at values ranging between $0 \le -t \le 3 {\rm ~GeV}^2$ for the pion
masses $m_\pi=930$, $760$, and $550 {\rm ~MeV}$, then extrapolated to
the physical pion mass. The corresponding nucleon masses are
$M_N=1724$, $1519$, and $1308$~MeV.  The $A$, $B$, and $C$ form
factors have been parametrized in Ref.~\cite{Gockeler:2003jfa} by a
dipole form with the mass of $M=1.1(2) {\rm ~GeV}$.\footnote{While in
Ref.~\cite{Gockeler:2003jfa} it is claimed that data disfavor a
monopole shape, it is also said that the results support the
assumption of the tensor meson dominance of the monopole form for
the spin-2 form factor.}

Note that in the presence of a heavy mass of a tensor meson the data
for the spin-0 gravitational form factor become noisy at large values
of the momentum transfer $t$ due to the relative smallness of the
signal and the propagation of errors in the formula for $\Theta_N^q$,
Eq.~(\ref{eq:thetaNq}).  For this reason we proceed as follows: with
the provided dipole fits we reconstruct the data points for the
spin-0 gravitational form factor using Eq.~(\ref{eq:thetaNq}) at the
quoted values of $t$, adding the statistical errors in quadrature. The
resulting $\Theta_N^q (t)$ becomes negative within errors at $-t>1.5
{\rm ~GeV}^2$, precluding accurate fitting with the monopole
form. Thus we determine the mass from the 
{\em slope} of the linearly-extrapolated form factor at $t=0$, 
according to the formula
$1/m_\sigma^2=\frac{d}{dt} \Theta_N^q(0)/ \Theta_N^q(0)$.  Figure
\ref{fig:sf_ms_mpi} shows the outcome of this procedure for the
available pion masses as well as for the data set resulting from
an extrapolation to the physical point made in Ref.~\cite{Gockeler:2003jfa}.

The result of the fit of the form (\ref{msmpi}) yields
\begin{eqnarray}
\ m_\sigma = 600^{+80}_{-100} {\rm MeV} , \;\;\; c=0.8(2), \label{best:sf-ms-c}
\end{eqnarray}
or  $m_\sigma^2 = 0.35{\rm ~GeV}^2 + 0.8 m_\pi^2 $.
These values are consistent within the uncertainties with the formula of the LHPC collaboration, $m_\sigma^2 = 0.387{\rm ~GeV}^2 + 0.91 m_\pi^2 $. 

Similarly, fitting to the form factors at the origin, we obtain  
\begin{eqnarray}
\langle x \rangle_{u+d}^N = 0.55(3), \;\;\; d=0.07(4). \label{xx}
\end{eqnarray}

\subsection{Summary of the $\sigma$ mass}

\begin{figure}
\begin{centering}
\includegraphics[width=.47\textwidth]{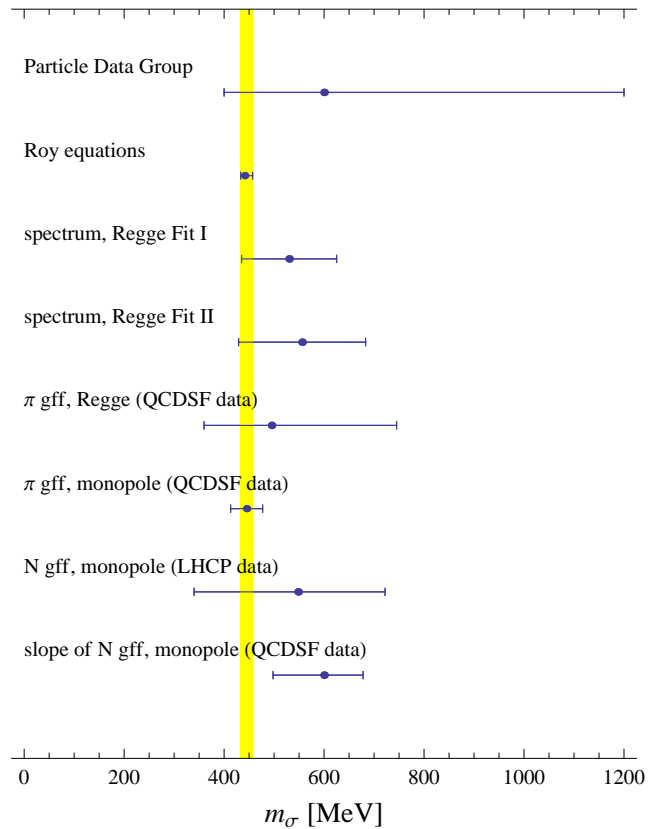} 
\end{centering}
\caption{(Color online) The collection of various results for the mass of the $sigma$
state. The vertical band indicates the ``benchmark'' calculation of
 Ref.~\cite{Caprini:2005zr}. The fits leading to the six bottom
 points are described in the text.
\label{fig:comp}}
\end{figure}

The various estimates of the mass of the lightest scalar-isoscalar
state, $m_\sigma$, are summarized in Fig.~\ref{fig:comp}. We note that
the values are compatible within the uncertainties. We remark that the
definition of the resonance mass via the pole position on the second
Riemann sheet of the scattering amplitude or via the Breit-Wigner
function may differ significantly, even by a factor of two. However, the difference occurs at
the level $1/N_c^2$ \cite{Nieves:2009kh}.

The previous results are based on extrapolation to physical pion
masses, a subject which has received much attention in recent years,
as chiral logs are expected to play a role. Comparison of
Figs.~\ref{fig:ms_mpi} and \ref{fig:sf_ms_mpi} provides a coherent
picture that the scalar mass grows with the current quark
mass. Actually, by analyzing the heavy pion limit, a distinction
between glueballs and mesons could be made. Indeed, glueballs are
existing states in pure gluodynamics which corresponds to the limit of
extremely heavy quarks.  The lightest $0^{++}$ glueball in that limit has a mass of
$M_{0^{++}}= 1.7 {\rm GeV}$, and this would correspond to a physical
state whose mass becomes a fixed number for infinitely heavy
quarks. Clearly, states which become $q \bar q $ states should
decouple since they have masses scaling as $2 m_q$ at large $m_q$. The
data are just too scarce and noisy to make a conclusive statement in that 
regard. From
this viewpoint, lattice data in the heavy-pion regime would be most
welcome.

\section{Calculations with a finite number of states \label{sec:more_states}}

In this Section we return to the derived sum rules in an attempt to
asses the nature of the scalar-isoscalar states.  The sum
rules~(\ref{epsilon},\ref{gtc},\ref{rel1},\ref{sr:0}) are:
\begin{eqnarray}
\sum_n f_n  g_{n\pi\pi}&=&1, \label{sr:repeat}\\ 
\sum_n f_n g_{n\pi\pi}m_n^2&=&0, \nonumber \\
\sum_n f_n  g_{nNN}&=&M_N, \nonumber \\ 
-\frac{1}{16} \sum_n f_n^2 m_n^2 &=& \epsilon_v. \nonumber  
\end{eqnarray}
In addition, the dimension-2 object of Eq.~(\ref{qcdsr},\ref{eq:sumfn^2}) is 
\begin{eqnarray}
\sum_n f_n^2  = C_2. \label{C2_gen}
\end{eqnarray}
It is clear that the satisfaction of the second sum rule
(\ref{sr:repeat}) with a finite number of states requires the presence of at least two
scalar-isoscalar states.  The presently-available lattice data do not
allow to extend the analysis of Sect.~\ref{sec:grav} to account for
two or more states.  Nevertheless, one can perform a qualitative
analysis based solely on Eq.~(\ref{sr:repeat}).

\begin{figure}
\begin{center}
\includegraphics[width=.47\textwidth]{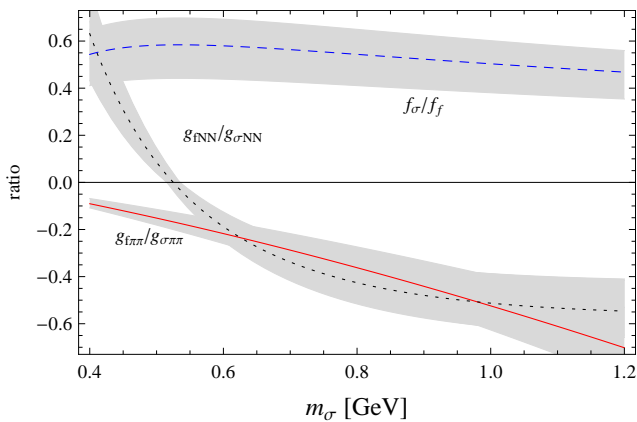} 
\end{center}
\caption{(Color online) Ratios of coupling constants of the $\sigma$ and $f_0(980)$
  in the two-state model. All ratios are of the order of
  $1/\sqrt{N_c}$, indicating the interpretation of the the $\sigma$ as
  a $q \bar q$ meson and $f_0(980)$ as a glueball, {\em cf.}
  Table~\ref{tab:nc}.
\label{fig:ratiog}}
\end{figure}

\begin{figure}
\begin{center}
\includegraphics[width=.47\textwidth]{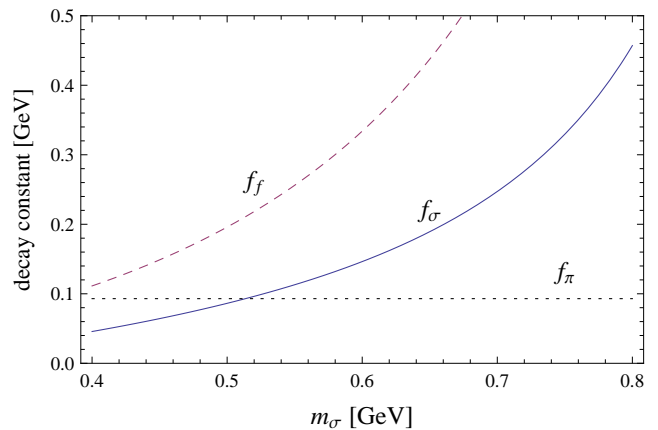} 
\end{center}
\caption{(Color online) The meson constants $f_n$ as functions of the $\sigma$
  mass.  \label{fig:fs_ff}}
\end{figure}

\begin{figure}
\begin{center}
\includegraphics[width=.47\textwidth]{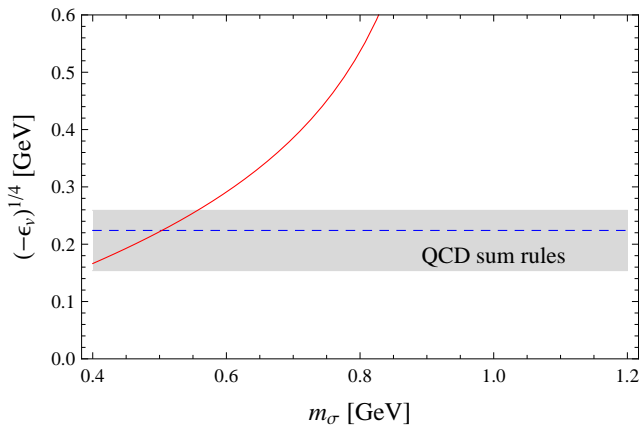} 
\end{center}
\caption{(Color online)
The fourth root of the minus energy density of the vacuum, $(-\epsilon_v)^{1/4}$ in the two-state model as a function of 
$m_\sigma$. The horizontal line and the band indicate the estimate with uncertainties of Eq.~(\ref{ioffe}) \cite{Ioffe:2002be}.
\label{fig:epsilon_v}}
\end{figure}

When we saturate the sum rules (\ref{sr:repeat}) with two
states, we get
\begin{eqnarray}
1 &=&g_{\sigma \pi \pi} f_\sigma + g_{f \pi \pi} f_f,  \label{two:states} \\
0 &=& g_{\sigma \pi \pi} f_\sigma m_\sigma^2 + g_{f \pi \pi} f_f m_f^2,  \\
M_N &=&  g_{\sigma NN } f_\sigma + g_{f NN} f_f, \label{third} \\ 
\epsilon_V &=& - \frac1{16} \left(f_\sigma^2 m_\sigma^2+ f_f^2 m_f^2 \right), \label{fourth} \\ 
C_2&=&f_\sigma^2 + f_f^2. \label{fifth} 
\end{eqnarray} 
Let us identify $\sigma$ with $f_0(600)$, the state $f$ with
$f_0(980)$, and treat $m_\sigma$ as a variable.  From the decay widths
$\Gamma_{\sigma\to \pi\pi}=540(20)$~MeV \cite{Caprini:2005zr} and
$\Gamma_{f_0\to \pi\pi}=70(30)$~MeV we infer (up to the sign) with the
help of Eq.~(\ref{decay:gamma}) the $m_\sigma$-dependent values of the
coupling constants $g_{\sigma \pi \pi}$ and $g_{f \pi \pi}$.  Then,
equations (\ref{two:states}) are used to find the constants $f_\sigma$
and $f_f$. We choose the convention where both of these constants are
positive.  Next, assuming $g_{\sigma NN }=9.5(5)$, we obtain $g_{f
  NN}$ from Eq.~(\ref{third}). Finally, from Eq.~(\ref{fourth}) we
compute the vacuum energy density.  The results are displayed in
Figs.~\ref{fig:ratiog}, \ref{fig:fs_ff}, and \ref{fig:epsilon_v}.

The most important qualitative feature can be inferred from the
analysis of the ratios of coupling constants.  Figure~\ref{fig:ratiog}
shows the ratios of the constants $f_\sigma/f_f$,
$g_{f\pi\pi}/g_{\sigma\pi\pi}$, and $g_{fNN}/g_{\sigma NN}$.  All
these ratios are of the order $1/\sqrt{N_c}\sim 0.6$ for the whole PDG
range of $m_\sigma$. According to the scaling of Table~\ref{tab:nc},
this supports the view that {\em $\sigma$ is a $q \bar q$ meson, while
  $f_0(980)$ is a glueball}. With this assignment the results of
Fig.~\ref{fig:ratiog} emerge naturally.  Certainly, the conclusion
relies on the assumption of having just two states dominating the sum
rules. Yet, it is appealing in its simplicity. 
In the following section we will see that if infinitely many states are considered this
conclusion would not necessarily follow when a further sum rule (the vanishing of a
dimension-2 condensate) is imposed. 
Note that in the two-state model the dimension-2 object, $C_2$, is 
necessarily positive, see Eq.~(\ref{fifth}).

Figure~\ref{fig:fs_ff} shows the dependence of the constants
$f_\sigma$ and $f_f$ on $m_\sigma$. For comparison, we also plot the
pion decay constant $f_\pi=93$~MeV. We note that $f_\sigma$ is of the
order $f_\pi$ near $m_\sigma \sim 500$~MeV. As discussed in
Appendix~\ref{sec:lin-sig}, $f_\pi$ and $f_\sigma$ need not be equal
even if we postulate the scalar to be the chiral partner of the pion.

Finally, we examine the last sum rule (\ref{fourth}), concerning the
energy density of the vacuum due to the gluon condensate.  According
to Ref.~\cite{Ioffe:2002be}, we have (for three active flavors)
\begin{eqnarray}
\epsilon_v =- \frac{9}{32} \langle \frac{\alpha}{\pi} G^2 \rangle= -  (224^{+35}_{-70} {\rm ~MeV})^4. \label{ioffe}
\end{eqnarray}
We note that the two-state model favors the lower values of
$m_\sigma$, up to $550$~MeV, where the sum rule (\ref{fourth}) is
satisfied within the values (\ref{ioffe}).

\section{Dimension two condensates in the scalar sector \label{sec:dim2}}

In the previous section we have explored phenomenological consequences
of saturating the sum rules in the large-$N_c$ limit with a finite
number of states (two states). Such an approach is justified for observables where
the coupling constants to higher-mass states are sharply suppressed,
and we are essentially left with the dominance of the low-mass states. This
is, however, not always the case. For instance, as discussed in
Sect.~\ref{subsec:two-point}, proper matching of the two-point
functions to QCD necessarily requires infinitely many intermediate
states present in the correlator. Only that way the leading
logarithmic behavior (\ref{eq:C0}) can be reproduced. Similar studies
have been carried out in the past for other correlators in the Regge
approach~\cite{Golterman:2001nk,Beane:2001uj,Beane:2001em,Simonov:2001di,Golterman:2002mi,Afonin:2003gp,Afonin:2004yb,
RuizArriola:2006gq,Arriola:2006ii,Arriola:2006sv,Afonin:2007mj}.
Here we focus on the $\Pi_{\Theta \Theta}$ correlator defined in
Eq.~(\ref{PiTT}).

In QCD, keeping the leading terms in $\alpha$, this object has the
explicit twist expansion of the form~\cite{
  Novikov:1979va,Novikov:1981xj,Pascual:1982bv,Dominguez:1986td,Narison:1996fm}
\begin{eqnarray}
\Pi_{\Theta\Theta} &=& - \frac{\alpha^2 (N_c^2-1) \beta_0^2}{32 \pi^4}  Q^4 \log \frac{Q^2}{\mu^2} + C_2 Q^2 \nonumber \\
&+& \frac{\alpha \beta_0^2}{16 \pi} \langle \frac{\alpha}{\pi} G^2 \rangle + \dots \label{twist}
\end{eqnarray}
(note a factor of 16 difference between our definition (\ref{PiTT}) and Ref.~\cite{Narison:1996fm}, carrying over to the definition 
of $C_2$).

The dimension-2 gluon condensate, originally proposed by Celenza and
Shakin~\cite{Celenza:1986th}, appears as an elusive gauge-invariant
non-perturbative and non-local operator which generates the lowest
$1/Q^2$ power corrections in the twist expansion.  Its dynamical
origin remains unclear~\cite{Zakharov:2005cg,Narison:2005hb}, despite
certain evidence provided by the instanton model~\cite{Hutter:1993sc},
phenomenological QCD sum rules reanalyses~\cite{Chetyrkin:1998yr},
phenomenological studies of the $\tau$ decay
data~\cite{Dominguez:1994qt}, further theoretical
considerations~\cite{Gubarev:2000eu,Gubarev:2000nz,Kondo:2001nq,Verschelde:2001ia},
quark-model calculations~\cite{Dorokhov:2003kf,Dorokhov:2006ac}, or
lattice simulations~\cite{Boucaud:2001st,RuizArriola:2004en} and their
relevance for the confinement-deconfinement phase
transition~\cite{Megias:2005ve}.  Phenomenologically, upper bounds for
the dimension-2 operators coming from the $e^+ e^-$ data were analyzed
in \cite{Narison:1992ru,Dominguez:1999xa,Dominguez:2006ct}, yielding
$16 C_2 \le 0.1~{\rm GeV}^2\sim (0.3~{\rm GeV})^2 $.  The possible
appearance of the dimension-2 condensates in Regge-like models can be
traced in Ref.~\cite{Shifman:2000jv}; it was also briefly discussed in
Ref.~\cite{Afonin:2004yb}. A more quantitative analysis was carried
out by us in Refs.~\cite{RuizArriola:2006gq,Arriola:2006sv}.

Very recently, the presence of the dimension-2 objects has been linked
to the truncation of the full perturbative series
\cite{Narison:2009ag}, indicating duality between the resummed
perturbation theory and the presence of the lowest power
corrections. This is supported by the correlation between perturbative
and non-perturbative contributions found in Ref.~\cite{Megias:2009ar}.

As we have already discussed in Sect.~\ref{subsec:two-point}, in the
$\Theta$-$\Theta$ correlator the dimension-2 object, $C_2 = \sum_n
f_n^2 $, appears naturally. Clearly, if there is a finite number of
states, the sum is necessarily positive and hence non-vanishing. This
is in an open contradiction to the Operator-Product-Expansion
philosophy, where only dimension-4 and higher {\it local}
gauge-invariant operators are allowed.  Therefore, the only scenario
where the dimension-2 object can possibly vanish is by assuming an
infinite number of resonances, such that the sum diverges requiring
regularization. In this case the asymptotic condition,
Eq.~(\ref{eq:f-asymp}), holds. As usual in the Regge models, we assume
$f_{n}$ to be independent on $n$ for all values of $n$, not only
asymptotically, as requested by Eq.~(\ref{eq:massS}).

The basic object of our study is the Regge sum (in this section $M$ denotes the lowest glueball mass)
\begin{eqnarray}
\Pi_{\Theta \Theta}=\sum_n \frac{Q^4 f_n^2}{m_n^2 + Q^2} = \sum_{n=0}^\infty \frac{Q^4 f_n^2}{M^2 + a n + Q^2}, \label{basic:Regge}
\end{eqnarray}
which can be evaluated using the formula (see Appendix~\ref{sec:zeta})
\begin{eqnarray}
&&\sum_{n=0}^\infty \left[\frac1{M^2 + a n +Q^2}- \frac1{M^2 + a n}
\right] \nonumber \\ &=&
\frac1{a} \left[ \Psi \left( \frac{M^2}{a}\right)- \Psi
\left( \frac{M^2+Q^2}{a}\right) \right] , 
\end{eqnarray}
where 
\begin{eqnarray}
\Psi(z) &=& \Gamma'(z)/\Gamma(z) \nonumber \\ &=& \log z - \frac1{2z} -
\frac1{12 z^2} + \frac1{120 z^4}+ {\cal O} (z^{-6}) \, ,
\end{eqnarray}
is the polygamma function.

We note that the infinite Regge sum generates properly the $Q^4 \log Q^2$
term. We assume that we have one family of states contributing to the Regge
sum at the leading-$N_c$ order. The matching with the leading term of
the OPE expansion gives immediately
\begin{eqnarray}
C_0=-\frac{f^2}{a} = -  \frac{\alpha^2 (N_c^2-1) \beta_0^2}{32 \pi^4}. \label{match:C0}
\end{eqnarray}
We read off that $f \sim N_c$, in accordance with the fact that the
glueballs provide the saturation at the leading-$N_c$
level. Explicitly,
\begin{eqnarray}
f= \frac{11 \sqrt{a} \alpha N_c^2}{12 \sqrt{2} \pi ^2}, \label{match:f}
\end{eqnarray}
which for $\alpha=0.2$, $a=1.31 {\rm GeV}^2$, and $N_c=3$ gives $f
\sim 135$~MeV, a reasonable value.  Note that taking the full slope
$a=1.31$~GeV is equivalent to taking every second state in the
scalar-isoscalar trajectory, hence we are implicitly assuming that
every second state on the trajectory is a glueball, alternating with
$q \bar q$ scalar-isoscalar mesons.  This alternation is suggested by
the two parallel trajectories, Eq.~(\ref{twotraj}).

For the case of the dimension-2 condensate the $\zeta$-function
regularization \cite{Arriola:2006sv} (see Appendix~\ref{sec:zeta}) 
applied on the spectrum $M_n^2 = a n + M^2$ yields  
\begin{eqnarray}
C_2 = \sum_n f_n^2 = f^2 \left( \frac12 - \frac{M^2}{a} \right). \label{match:dim2}  
\end{eqnarray}
The condition for a vanishing dimension-2 condensate is thus equivalent to
requesting that the mass of lowest state be given by the equality
\begin{eqnarray}
M^2 = \frac12 a, 
\end{eqnarray} 
which for $a=1.31(12)$~GeV from Eq.~(\ref{fitII}) means $M = 810(40)
{\rm ~MeV}$. The value is not far from the mass of $f_0(980)$, it is
also in the PDG range for $m_\sigma$
(cf.~Fig.~\ref{fig:comp}).\footnote{As described
  in~\cite{Nieves:2009kh}, the shift from the pole resonance to the
  Breit-Wigner resonance is ${\cal O} (N_c^{-2})$ suppressed,
  suggesting that the leading $N_c$ values lies in between.}  Thus, on
the basis of a simplest Regge model considered here it is not possible
to sort out which state is a glueball, and which one is a $q \bar q$
meson, if the dimension-2 condensate were indeed to
vanish.

{The condition for the vanishing $C_2$ is identical to
  that for the vector or axial flavor channels with $C_{2,V}=\sum_n
  f_{n,V}^2$ and $C_{2,A}=\sum_n f_{n,A}^2$, 
  respectively~\cite{RuizArriola:2006gq}. Note that in that case the
  first Weinberg sum rule requires a non-vanishing dimension-2 object
  $f_\pi^2 = C_{2,A}- C_{2,V}$. This shows that both dimension-2 operators
  cannot vanish simultaneously, despite the fact that they cannot be
  represented by local and gauge invariant operators.}

One may extend the Regge model by modifying the parameters of the
lowest states on the trajectory, for instance the $f$ constants.  It
is known from similar studies that such modifications cause strong
effects~\cite{RuizArriola:2006gq}.  Also, the $1/N_c$ effects may play
a role. Nevertheless, we see that the fine-tuning needed to cancel the
dimension-2 object leads to a quite natural condition for the lowest
mass of the Regge family.

\begin{figure}
\begin{center}
\includegraphics[width=.47\textwidth]{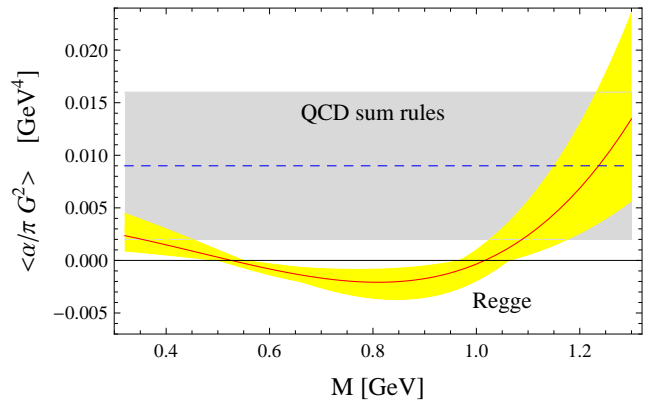} 
\end{center}
\caption{(Color online)
The gluon condensate, $\langle \frac{\alpha}{\pi} G^2\rangle$, in the one-family Regge model plotted as a function of 
the mass of the lowest state in the glueball trajectory, $M$. We use $a=1.31(12)$~GeV and $\alpha=0.2(1)$. The band indicates the 
uncertainty in these quantities. The horizontal dashed line and the band indicate the estimate of Ref.~\cite{Ioffe:2002be}, equal to 
$0.009(7)$~GeV${}^4$.
\label{fig:g4}}
\end{figure}

For the dimension-4 condensate the matching of Eq.~(\ref{basic:Regge}) to Eq.~(\ref{twist}) leads to the 
large-$N_c$ identity (see Appendix~\ref{sec:zeta})
\begin{eqnarray}
\frac{121 N_c^2 \alpha}{144 \pi} \langle \frac{\alpha}{\pi} G^2\rangle &=& -\frac12 \sum_n f_n^2 M_n^2 \nonumber \\ 
&=& \frac{f^2}{2} 
\left ( \frac{M^4}{a}+\frac{a}{6}-M^2 \right ). \label{match:dim4}   
\end{eqnarray}
The condition for the gluon condensate to be positive is therefore $a > (3+\sqrt{3})M^2$ or $ a < (3-\sqrt{3}) M^2$.
Combining Eq.~(\ref{match:dim4}) and (\ref{match:f}) yields 
\begin{eqnarray}
\langle \frac{\alpha}{\pi} G^2\rangle = \frac{\alpha N_c^2 \left(a^2-6 a M^2+6 M^4\right)}{24 \pi^3}. \label{m:g4}
\end{eqnarray}
This quantity is plotted against the mass $M$ in Fig.~\ref{fig:g4}. We
note that we have two windows for the values of $M$, above 1.3~GeV and
below 0.45~GeV, which match the predictions of the considered Regge
model to the QCD estimate of Ref.~\cite{Ioffe:2002be} giving $\langle
\frac{\alpha}{\pi} G^2\rangle= 0.009(7)~{\rm GeV}^4$. Alternative
determinations yield about twice larger values, $\langle
\frac{\alpha}{\pi} G^2\rangle= 0.022(3)~{\rm
  GeV}^4$~\cite{Narison:1995tw}, or even negative values, $\langle
\frac{\alpha}{\pi} G^2\rangle= -0.005(1)~{\rm
  GeV}^4$~\cite{Davier:2008sk}. In our model the gluon condensate is
related to the lighest glueball mass $M$ accorging to Eq.~(\ref{match:dim4}). Modifications of the Regge model along the
lines of Ref.~\cite{RuizArriola:2006gq,Arriola:2006sv} may modify or
lift these constraints, making it possible for the approach to be
fully consistent with the requirements of the Operator Product
Expansion.

\section{Conclusions}
\label{sec:concl}

We have presented a wide-scope analysis of the $\sigma$-state in the
large-$N_c$ Regge framework.  Our main points are the following:
\begin{enumerate}
\item The $\sigma$ state fits very well to the scalar-isoscalar Regge
  trajectory including {\it all} the known states from the PDG
  review. From the viewpoint of the spectrum, there seems to be no
  particular distinction among the scalar-isoscalar states. The {\em
    glueball} or {\em meson} nature has to be sought based on the
  $N_c$ scaling of the couplings of the $0^{++}$ states to the vacuum or
  to hadrons.
 
\item The mass of the lowest $0^{++}$ state (the $\sigma$) can also be
  extracted from the analysis of gravitational form factors of the
  pion and nucleon, available from the lattice
  calculations. Extracted masses are compatible with the benchmark
  determinations within the expected accuracy of the large-$N_c$
  framework, showing the consistency of the approach and the proximity
  of the large-$N_c$ limit to the real world in the scalar-isoscalar
  sector.

\item The constraints of chiral symmetry and the high-energy behavior
  of the energy-momentum correlator imposes a set of sum rules
  involving various coupling constants. Saturation of these sum rules
  with just two states suggest that $f_0 (600)$ is a meson and
  $f_0(980)$ is a glueball. Such an assignment, specific to the simple
  two-state model, results in natural $N_c$-scaling of the ratios of
  the coupling constants of these states.

\item Correlation functions of the energy-momentum tensor are
  saturated by (infinitely many) scalar-isoscalar states. The analysis
  of the Operator Product Expansion shows that there appears the
  dimension-2 condensate, whose vanishing requires a fine tuning of
  parameters. Then, in the simplest Regge model the resulting mass of
  the lowest glueball state is $\sim 800$~MeV.

\end{enumerate}

We thank J. Prades for clarifications and S. Narison for remarks.
This work is partially supported by the Polish Ministry of Science and
Higher Education (grants N202~034~32/0918 and N202~249235), Spanish
DGI and FEDER funds (grant FIS2008-01143/FIS), Junta de Andaluc{\'\i}a
(grant FQM225-05), the EU Integrated Infrastructure Initiative Hadron
Physics Project (contract RII3-CT-2004-506078).

\appendix

\section{The $\zeta$-regularization \label{sec:zeta}}

In the large $N_c$ limit only tree-level diagrams survive, however,
with infinitely many resonances. Thus, the infinite sums may need to
be regularized. In the dimensional regularization the coupling of the
resonance to the current acquires an additional dimension $f_n \to f_n
\mu^{\epsilon}$ with $\epsilon = d-4$. By choosing as a natural scale
$\mu=m_n$, one gets $ f_n^2 \to f_n^2 m_n^{2 \epsilon}$, which means  
\begin{eqnarray}
\sum_n f_n^2 m_n^{2k} \equiv \lim_{s \to k} \sum_n f_n^2
m_n^{2 s} .
\label{eq:zeta}
\end{eqnarray}
This is completely equivalent to formally expanding the
two-point correlator  at large $Q^2$ and reinterpreting the coefficients within the
zeta-function regularization~\cite{Arriola:2006sv}. In our case the
mass formula is $m_n^2 = a n + m_0^2$, hence one has to consider the
sums of the form
\begin{eqnarray}
\sum_n (n + \alpha)^{2k} \equiv \lim_{s \to k} \sum_n 
(n +\alpha)^{2 s} = \zeta (-2 s , \alpha),
\end{eqnarray}
which is the generalized Riemann zeta function. Then, we arrive at the
explicit formulas
\begin{eqnarray}
\sum_{n=0}^\infty (n+\alpha)^{0} &=& \frac12 - \alpha, \\
\sum_{n=0}^\infty (n+\alpha)^{1} &=& \frac12 \left(-\alpha^2 + \alpha- \frac16 \right), \\ 
\sum_{n=0}^\infty (n+\alpha)^{2} &=& \frac13 \left( - \alpha^3 + \frac32 \alpha^2 - \frac{\alpha}{2}\right). 
\end{eqnarray}

\section{The linear Lagrangean for one scalar \label{sec:lin-sig}}

For only one scalar-isoscalar state, Lagrangean (\ref{lagpi}) with condition 
(\ref{eq:cond1}) reads
\begin{eqnarray}
{\cal L}_{\Phi,\pi} &=& \frac{f_\pi^2}{4 f_\sigma^2} \Phi^2 \langle
\partial^\mu U^\dagger \partial_\mu U^\dagger \rangle + \frac12
\partial^\mu \Phi \partial_\mu \Phi - V(\Phi) . \nonumber \\
\end{eqnarray}
Introducing the Cartesian decomposition
\begin{eqnarray}
 \Sigma = \frac{f_\pi}{f_\sigma} \Phi U = \sigma + i \vec \tau \cdot \vec \pi,
\end{eqnarray}
one gets an extension of the linear sigma model with 
the so-called $A$-term \cite{Broniowski:2002ew}, allowed in the linear realization of the chiral 
symmetry:
\begin{eqnarray}
{\cal L} &=& \frac{1}{2} \left[ \partial^\mu \sigma \partial_\mu
\sigma + \partial^\mu \vec \pi \cdot \partial_\mu \vec \pi \right]
+ \frac{A^2}{2} \left[ \sigma \partial^\mu \sigma + \vec \pi \cdot \partial_\mu \vec \pi \right]^2
\nonumber \\ &-& V(\sigma^2+\vec \pi^2), \label{A_term}
\end{eqnarray}
where in the present case
\begin{eqnarray}
A^2 \sigma_{\rm vac}^2=\left (\frac{f_\sigma^2}{f_\pi^2}-1 \right).
\end{eqnarray}
Note that $\sigma$ and $\vec \pi$ are chiral partners.

Imposing the condition $f_\sigma=f_\pi$ removes the $A$ term. 
Note, however,  that the identity $f_\pi = f_\sigma$ is not a
consequence of any symmetry or physical requirement. With 
this condition the Lagrangean reads 
\begin{eqnarray}
{\cal L} &=& \frac{1}{2} \left[ \partial^\mu \sigma \partial_\mu
\sigma + \partial^\mu \vec \pi \cdot \partial_\mu \vec \pi \right]
\nonumber \\ &-& \frac{m_\sigma^2}{8 f_\pi^2} \left(\sigma^2 + \vec
\pi^2\right)^2 \left[ \log \left(\frac{\sigma^2 + \vec
\pi^2}{f_\pi^2}\right) - \frac12 \right]. \label{pot:sp}
\end{eqnarray}
Note the resemblance to the bosonic part of the dilated chiral
quark model of Ref.~\cite{Beane:1994ds}. The main difference between
potential (\ref{pot:sp}) and the standard Mexican hat potential,
\begin{eqnarray}
V(\sigma , \vec \pi) =  \frac{m_\sigma^2}{8 f_\pi^2} \left(\sigma^2 + 
\vec \pi^2- f_\pi^2 \right)^2, 
\end{eqnarray}
is the scale behaviour of the latter, a different vacuum energy
density (factor of two difference), and modified couplings containing
more than one scalar. The $\sigma \pi \pi $ vertex is the same {\it on
  shell}, thus yields the same decay width of the $\sigma$ into two
pions.


\begin{thebibliography}{110}
\expandafter\ifx\csname natexlab\endcsname\relax\def\natexlab#1{#1}\fi
\expandafter\ifx\csname bibnamefont\endcsname\relax
  \def\bibnamefont#1{#1}\fi
\expandafter\ifx\csname bibfnamefont\endcsname\relax
  \def\bibfnamefont#1{#1}\fi
\expandafter\ifx\csname citenamefont\endcsname\relax
  \def\citenamefont#1{#1}\fi
\expandafter\ifx\csname url\endcsname\relax
  \def\url#1{\texttt{#1}}\fi
\expandafter\ifx\csname urlprefix\endcsname\relax\def\urlprefix{URL }\fi
\providecommand{\bibinfo}[2]{#2}
\providecommand{\eprint}[2][]{\url{#2}}

\bibitem[{\citenamefont{Klempt and Zaitsev}(2007)}]{Klempt:2007cp}
\bibinfo{author}{\bibfnamefont{E.}~\bibnamefont{Klempt}} \bibnamefont{and}
  \bibinfo{author}{\bibfnamefont{A.}~\bibnamefont{Zaitsev}},
  \bibinfo{journal}{Phys. Rept.} \textbf{\bibinfo{volume}{454}},
  \bibinfo{pages}{1} (\bibinfo{year}{2007}), \eprint{0708.4016}.

\bibitem[{\citenamefont{{Rupp} et~al.}(2008)\citenamefont{{Rupp}, {van
  Beveren}, {Bicudo}, {Hiller}, and {Kleefeld}}}]{2008AIPC.1030.....R}
\bibinfo{editor}{\bibfnamefont{G.}~\bibnamefont{{Rupp}}},
  \bibinfo{editor}{\bibfnamefont{E.}~\bibnamefont{{van Beveren}}},
  \bibinfo{editor}{\bibfnamefont{P.}~\bibnamefont{{Bicudo}}},
  \bibinfo{editor}{\bibfnamefont{B.}~\bibnamefont{{Hiller}}}, \bibnamefont{and}
  \bibinfo{editor}{\bibfnamefont{F.}~\bibnamefont{{Kleefeld}}}, eds.,
  \emph{\bibinfo{title}{{SCADRON70: Workshop on Scalar Mesons and Related
  Topics Honoring Michael Scadron's 70th Birthday}}}, vol.
  \bibinfo{volume}{1030} of \emph{\bibinfo{series}{American Institute of
  Physics Conference Series}} (\bibinfo{year}{2008}).

\bibitem[{\citenamefont{Achasov and Shestakov}(2009)}]{Achasov:2009ee}
\bibinfo{author}{\bibfnamefont{N.~N.} \bibnamefont{Achasov}} \bibnamefont{and}
  \bibinfo{author}{\bibfnamefont{G.~N.} \bibnamefont{Shestakov}}
  (\bibinfo{year}{2009}), \eprint{0905.2017}.

\bibitem[{\citenamefont{Johnson and Teller}(1955)}]{PhysRev.98.783}
\bibinfo{author}{\bibfnamefont{M.~H.} \bibnamefont{Johnson}} \bibnamefont{and}
  \bibinfo{author}{\bibfnamefont{E.}~\bibnamefont{Teller}},
  \bibinfo{journal}{Phys. Rev.} \textbf{\bibinfo{volume}{98}},
  \bibinfo{pages}{783} (\bibinfo{year}{1955}).

\bibitem[{\citenamefont{Machleidt}(2001)}]{Machleidt:2000ge}
\bibinfo{author}{\bibfnamefont{R.}~\bibnamefont{Machleidt}},
  \bibinfo{journal}{Phys. Rev.} \textbf{\bibinfo{volume}{C63}},
  \bibinfo{pages}{024001} (\bibinfo{year}{2001}), \eprint{nucl-th/0006014}.

\bibitem[{\citenamefont{Gell-Mann and Levy}(1960)}]{GellMann:1960np}
\bibinfo{author}{\bibfnamefont{M.}~\bibnamefont{Gell-Mann}} \bibnamefont{and}
  \bibinfo{author}{\bibfnamefont{M.}~\bibnamefont{Levy}},
  \bibinfo{journal}{Nuovo Cim.} \textbf{\bibinfo{volume}{16}},
  \bibinfo{pages}{705} (\bibinfo{year}{1960}).

\bibitem[{\citenamefont{Weinberg}(1968)}]{Weinberg:1968de}
\bibinfo{author}{\bibfnamefont{S.}~\bibnamefont{Weinberg}},
  \bibinfo{journal}{Phys. Rev.} \textbf{\bibinfo{volume}{166}},
  \bibinfo{pages}{1568} (\bibinfo{year}{1968}).

\bibitem[{\citenamefont{Gasser and Leutwyler}(1984)}]{Gasser:1983yg}
\bibinfo{author}{\bibfnamefont{J.}~\bibnamefont{Gasser}} \bibnamefont{and}
  \bibinfo{author}{\bibfnamefont{H.}~\bibnamefont{Leutwyler}},
  \bibinfo{journal}{Ann. Phys.} \textbf{\bibinfo{volume}{158}},
  \bibinfo{pages}{142} (\bibinfo{year}{1984}).

\bibitem[{\citenamefont{Tornqvist and Roos}(1996)}]{Tornqvist:1995ay}
\bibinfo{author}{\bibfnamefont{N.~A.} \bibnamefont{Tornqvist}}
  \bibnamefont{and} \bibinfo{author}{\bibfnamefont{M.}~\bibnamefont{Roos}},
  \bibinfo{journal}{Phys. Rev. Lett.} \textbf{\bibinfo{volume}{76}},
  \bibinfo{pages}{1575} (\bibinfo{year}{1996}), \eprint{hep-ph/9511210}.

\bibitem[{\citenamefont{Yao et~al.}(2006)}]{Yao:2006px}
\bibinfo{author}{\bibfnamefont{W.~M.} \bibnamefont{Yao}} \bibnamefont{et~al.}
  (\bibinfo{collaboration}{Particle Data Group}), \bibinfo{journal}{J. Phys.}
  \textbf{\bibinfo{volume}{G33}}, \bibinfo{pages}{1} (\bibinfo{year}{2006}).

\bibitem[{\citenamefont{{van Beveren} et~al.}(2002)\citenamefont{{van Beveren},
  Kleefeld, Rupp, and Scadron}}]{vanBeveren:2002mc}
\bibinfo{author}{\bibfnamefont{E.}~\bibnamefont{{van Beveren}}},
  \bibinfo{author}{\bibfnamefont{F.}~\bibnamefont{Kleefeld}},
  \bibinfo{author}{\bibfnamefont{G.}~\bibnamefont{Rupp}}, \bibnamefont{and}
  \bibinfo{author}{\bibfnamefont{M.~D.} \bibnamefont{Scadron}},
  \bibinfo{journal}{Mod. Phys. Lett.} \textbf{\bibinfo{volume}{A17}},
  \bibinfo{pages}{1673} (\bibinfo{year}{2002}), \eprint{hep-ph/0204139}.

\bibitem[{\citenamefont{Caprini et~al.}(2006)\citenamefont{Caprini, Colangelo,
  and Leutwyler}}]{Caprini:2005zr}
\bibinfo{author}{\bibfnamefont{I.}~\bibnamefont{Caprini}},
  \bibinfo{author}{\bibfnamefont{G.}~\bibnamefont{Colangelo}},
  \bibnamefont{and}
  \bibinfo{author}{\bibfnamefont{H.}~\bibnamefont{Leutwyler}},
  \bibinfo{journal}{Phys. Rev. Lett.} \textbf{\bibinfo{volume}{96}},
  \bibinfo{pages}{132001} (\bibinfo{year}{2006}), \eprint{hep-ph/0512364}.

\bibitem[{\citenamefont{Kaminski et~al.}(2008)\citenamefont{Kaminski, Pelaez,
  and Yndurain}}]{Kaminski:2006qe}
\bibinfo{author}{\bibfnamefont{R.}~\bibnamefont{Kaminski}},
  \bibinfo{author}{\bibfnamefont{J.~R.} \bibnamefont{Pelaez}},
  \bibnamefont{and} \bibinfo{author}{\bibfnamefont{F.~J.}
  \bibnamefont{Yndurain}}, \bibinfo{journal}{Phys. Rev.}
  \textbf{\bibinfo{volume}{D77}}, \bibinfo{pages}{054015}
  (\bibinfo{year}{2008}), \eprint{0710.1150}.

\bibitem[{\citenamefont{Nieves and Arriola}(2009)}]{Nieves:2009ez}
\bibinfo{author}{\bibfnamefont{J.}~\bibnamefont{Nieves}} \bibnamefont{and}
  \bibinfo{author}{\bibfnamefont{E.~R.} \bibnamefont{Arriola}},
  \bibinfo{journal}{Phys. Rev.} \textbf{\bibinfo{volume}{D80}},
  \bibinfo{pages}{045023} (\bibinfo{year}{2009}), \eprint{0904.4344}.

\bibitem[{\citenamefont{'t~Hooft et~al.}(2008)\citenamefont{'t~Hooft, Isidori,
  Maiani, Polosa, and Riquer}}]{Hooft:2008we}
\bibinfo{author}{\bibfnamefont{G.}~\bibnamefont{'t~Hooft}},
  \bibinfo{author}{\bibfnamefont{G.}~\bibnamefont{Isidori}},
  \bibinfo{author}{\bibfnamefont{L.}~\bibnamefont{Maiani}},
  \bibinfo{author}{\bibfnamefont{A.~D.} \bibnamefont{Polosa}},
  \bibnamefont{and} \bibinfo{author}{\bibfnamefont{V.}~\bibnamefont{Riquer}},
  \bibinfo{journal}{Phys. Lett.} \textbf{\bibinfo{volume}{B662}},
  \bibinfo{pages}{424} (\bibinfo{year}{2008}), \eprint{0801.2288}.

\bibitem[{\citenamefont{Fariborz et~al.}(2009)\citenamefont{Fariborz, Jora, and
  Schechter}}]{Fariborz:2009cq}
\bibinfo{author}{\bibfnamefont{A.~H.} \bibnamefont{Fariborz}},
  \bibinfo{author}{\bibfnamefont{R.}~\bibnamefont{Jora}}, \bibnamefont{and}
  \bibinfo{author}{\bibfnamefont{J.}~\bibnamefont{Schechter}},
  \bibinfo{journal}{Phys. Rev.} \textbf{\bibinfo{volume}{D79}},
  \bibinfo{pages}{074014} (\bibinfo{year}{2009}), \eprint{0902.2825}.

\bibitem[{\citenamefont{Kaminski et~al.}(2009)\citenamefont{Kaminski,
  Mennessier, and Narison}}]{Kaminski:2009qg}
\bibinfo{author}{\bibfnamefont{R.}~\bibnamefont{Kaminski}},
  \bibinfo{author}{\bibfnamefont{G.}~\bibnamefont{Mennessier}},
  \bibnamefont{and} \bibinfo{author}{\bibfnamefont{S.}~\bibnamefont{Narison}},
  \bibinfo{journal}{Phys. Lett.} \textbf{\bibinfo{volume}{B680}},
  \bibinfo{pages}{148} (\bibinfo{year}{2009}), \eprint{0904.2555}.

\bibitem[{\citenamefont{{'t Hooft}}(1974)}]{'tHooft:1973jz}
\bibinfo{author}{\bibfnamefont{G.}~\bibnamefont{{'t Hooft}}},
  \bibinfo{journal}{Nucl. Phys.} \textbf{\bibinfo{volume}{B72}},
  \bibinfo{pages}{461} (\bibinfo{year}{1974}).

\bibitem[{\citenamefont{Witten}(1979)}]{Witten:1979kh}
\bibinfo{author}{\bibfnamefont{E.}~\bibnamefont{Witten}},
  \bibinfo{journal}{Nucl. Phys.} \textbf{\bibinfo{volume}{B160}},
  \bibinfo{pages}{57} (\bibinfo{year}{1979}).

\bibitem[{\citenamefont{Teper}(2008)}]{Teper:2008yi}
\bibinfo{author}{\bibfnamefont{M.}~\bibnamefont{Teper}} (\bibinfo{year}{2008}),
  \eprint{0812.0085}.

\bibitem[{\citenamefont{Sannino and Schechter}(1995)}]{Sannino:1995ik}
\bibinfo{author}{\bibfnamefont{F.}~\bibnamefont{Sannino}} \bibnamefont{and}
  \bibinfo{author}{\bibfnamefont{J.}~\bibnamefont{Schechter}},
  \bibinfo{journal}{Phys. Rev.} \textbf{\bibinfo{volume}{D52}},
  \bibinfo{pages}{96} (\bibinfo{year}{1995}), \eprint{hep-ph/9501417}.

\bibitem[{\citenamefont{Pelaez}(2004)}]{Pelaez:2003dy}
\bibinfo{author}{\bibfnamefont{J.~R.} \bibnamefont{Pelaez}},
  \bibinfo{journal}{Phys. Rev. Lett.} \textbf{\bibinfo{volume}{92}},
  \bibinfo{pages}{102001} (\bibinfo{year}{2004}), \eprint{hep-ph/0309292}.

\bibitem[{\citenamefont{Pelaez and Rios}(2006)}]{Pelaez:2006nj}
\bibinfo{author}{\bibfnamefont{J.~R.} \bibnamefont{Pelaez}} \bibnamefont{and}
  \bibinfo{author}{\bibfnamefont{G.}~\bibnamefont{Rios}},
  \bibinfo{journal}{Phys. Rev. Lett.} \textbf{\bibinfo{volume}{97}},
  \bibinfo{pages}{242002} (\bibinfo{year}{2006}), \eprint{hep-ph/0610397}.

\bibitem[{\citenamefont{Nieves et~al.}(2002)\citenamefont{Nieves, {Pavon
  Valderrama}, and {Ruiz Arriola}}}]{Nieves:2001de}
\bibinfo{author}{\bibfnamefont{J.}~\bibnamefont{Nieves}},
  \bibinfo{author}{\bibfnamefont{M.}~\bibnamefont{{Pavon Valderrama}}},
  \bibnamefont{and} \bibinfo{author}{\bibfnamefont{E.}~\bibnamefont{{Ruiz
  Arriola}}}, \bibinfo{journal}{Phys. Rev.} \textbf{\bibinfo{volume}{D65}},
  \bibinfo{pages}{036002} (\bibinfo{year}{2002}), \eprint{hep-ph/0109077}.

\bibitem[{\citenamefont{Harada et~al.}(2004)\citenamefont{Harada, Sannino, and
  Schechter}}]{Harada:2003em}
\bibinfo{author}{\bibfnamefont{M.}~\bibnamefont{Harada}},
  \bibinfo{author}{\bibfnamefont{F.}~\bibnamefont{Sannino}}, \bibnamefont{and}
  \bibinfo{author}{\bibfnamefont{J.}~\bibnamefont{Schechter}},
  \bibinfo{journal}{Phys. Rev.} \textbf{\bibinfo{volume}{D69}},
  \bibinfo{pages}{034005} (\bibinfo{year}{2004}), \eprint{hep-ph/0309206}.

\bibitem[{\citenamefont{Nieves and {Ruiz Arriola}}(2000)}]{Nieves:1999bx}
\bibinfo{author}{\bibfnamefont{J.}~\bibnamefont{Nieves}} \bibnamefont{and}
  \bibinfo{author}{\bibfnamefont{E.}~\bibnamefont{{Ruiz Arriola}}},
  \bibinfo{journal}{Nucl. Phys.} \textbf{\bibinfo{volume}{A679}},
  \bibinfo{pages}{57} (\bibinfo{year}{2000}), \eprint{hep-ph/9907469}.

\bibitem[{\citenamefont{Calle~Cordon and
  Ruiz~Arriola}(2008)}]{CalleCordon:2008eu}
\bibinfo{author}{\bibfnamefont{A.}~\bibnamefont{Calle~Cordon}}
  \bibnamefont{and}
  \bibinfo{author}{\bibfnamefont{E.}~\bibnamefont{Ruiz~Arriola}},
  \bibinfo{journal}{AIP Conf. Proc.} \textbf{\bibinfo{volume}{1030}},
  \bibinfo{pages}{334} (\bibinfo{year}{2008}), \eprint{0804.2350}.

\bibitem[{\citenamefont{Kaplan and Savage}(1996)}]{Kaplan:1995yg}
\bibinfo{author}{\bibfnamefont{D.~B.} \bibnamefont{Kaplan}} \bibnamefont{and}
  \bibinfo{author}{\bibfnamefont{M.~J.} \bibnamefont{Savage}},
  \bibinfo{journal}{Phys. Lett.} \textbf{\bibinfo{volume}{B365}},
  \bibinfo{pages}{244} (\bibinfo{year}{1996}), \eprint{hep-ph/9509371}.

\bibitem[{\citenamefont{Kaplan and Manohar}(1997)}]{Kaplan:1996rk}
\bibinfo{author}{\bibfnamefont{D.~B.} \bibnamefont{Kaplan}} \bibnamefont{and}
  \bibinfo{author}{\bibfnamefont{A.~V.} \bibnamefont{Manohar}},
  \bibinfo{journal}{Phys. Rev.} \textbf{\bibinfo{volume}{C56}},
  \bibinfo{pages}{76} (\bibinfo{year}{1997}), \eprint{nucl-th/9612021}.

\bibitem[{\citenamefont{Banerjee et~al.}(2002)\citenamefont{Banerjee, Cohen,
  and Gelman}}]{Banerjee:2001js}
\bibinfo{author}{\bibfnamefont{M.~K.} \bibnamefont{Banerjee}},
  \bibinfo{author}{\bibfnamefont{T.~D.} \bibnamefont{Cohen}}, \bibnamefont{and}
  \bibinfo{author}{\bibfnamefont{B.~A.} \bibnamefont{Gelman}},
  \bibinfo{journal}{Phys. Rev.} \textbf{\bibinfo{volume}{C65}},
  \bibinfo{pages}{034011} (\bibinfo{year}{2002}), \eprint{hep-ph/0109274}.

\bibitem[{\citenamefont{Calle~Cordon and
  Ruiz~Arriola}(2009)}]{CalleCordon:2009ps}
\bibinfo{author}{\bibfnamefont{A.}~\bibnamefont{Calle~Cordon}}
  \bibnamefont{and}
  \bibinfo{author}{\bibfnamefont{E.}~\bibnamefont{Ruiz~Arriola}},
  \bibinfo{journal}{Phys. Rev.} \textbf{\bibinfo{volume}{C80}},
  \bibinfo{pages}{014002} (\bibinfo{year}{2009}), \eprint{0904.0421}.

\bibitem[{\citenamefont{Nieves and {Ruiz Arriola}}(2009)}]{Nieves:2009kh}
\bibinfo{author}{\bibfnamefont{J.}~\bibnamefont{Nieves}} \bibnamefont{and}
  \bibinfo{author}{\bibfnamefont{E.}~\bibnamefont{{Ruiz Arriola}}},
  \bibinfo{journal}{Phys. Lett.} \textbf{\bibinfo{volume}{B679}},
  \bibinfo{pages}{449} (\bibinfo{year}{2009}), \eprint{0904.4590}.

\bibitem[{\citenamefont{Corrigan and Ramond}(1979)}]{Corrigan:1979xf}
\bibinfo{author}{\bibfnamefont{E.}~\bibnamefont{Corrigan}} \bibnamefont{and}
  \bibinfo{author}{\bibfnamefont{P.}~\bibnamefont{Ramond}},
  \bibinfo{journal}{Phys. Lett.} \textbf{\bibinfo{volume}{B87}},
  \bibinfo{pages}{73} (\bibinfo{year}{1979}).

\bibitem[{\citenamefont{Kiritsis and Papavassiliou}(1990)}]{Kiritsis:1989ge}
\bibinfo{author}{\bibfnamefont{E.~B.} \bibnamefont{Kiritsis}} \bibnamefont{and}
  \bibinfo{author}{\bibfnamefont{J.}~\bibnamefont{Papavassiliou}},
  \bibinfo{journal}{Phys. Rev.} \textbf{\bibinfo{volume}{D42}},
  \bibinfo{pages}{4238} (\bibinfo{year}{1990}).

\bibitem[{\citenamefont{Sannino and Schechter}(2007)}]{Sannino:2007yp}
\bibinfo{author}{\bibfnamefont{F.}~\bibnamefont{Sannino}} \bibnamefont{and}
  \bibinfo{author}{\bibfnamefont{J.}~\bibnamefont{Schechter}},
  \bibinfo{journal}{Phys. Rev.} \textbf{\bibinfo{volume}{D76}},
  \bibinfo{pages}{014014} (\bibinfo{year}{2007}), \eprint{0704.0602}.

\bibitem[{\citenamefont{Amsler et~al.}(2008)}]{Amsler:2008zz}
\bibinfo{author}{\bibfnamefont{C.}~\bibnamefont{Amsler}} \bibnamefont{et~al.}
  (\bibinfo{collaboration}{Particle Data Group}), \bibinfo{journal}{Phys.
  Lett.} \textbf{\bibinfo{volume}{B667}}, \bibinfo{pages}{1}
  (\bibinfo{year}{2008}).

\bibitem[{\citenamefont{Anisovich et~al.}(2000)\citenamefont{Anisovich,
  Anisovich, and Sarantsev}}]{Anisovich:2000kxa}
\bibinfo{author}{\bibfnamefont{A.~V.} \bibnamefont{Anisovich}},
  \bibinfo{author}{\bibfnamefont{V.~V.} \bibnamefont{Anisovich}},
  \bibnamefont{and} \bibinfo{author}{\bibfnamefont{A.~V.}
  \bibnamefont{Sarantsev}}, \bibinfo{journal}{Phys. Rev.}
  \textbf{\bibinfo{volume}{D62}}, \bibinfo{pages}{051502}
  (\bibinfo{year}{2000}), \eprint{hep-ph/0003113}.

\bibitem[{\citenamefont{Anisovich}(2006)}]{Anisovich:2005jn}
\bibinfo{author}{\bibfnamefont{V.~V.} \bibnamefont{Anisovich}},
  \bibinfo{journal}{Int. J. Mod. Phys.} \textbf{\bibinfo{volume}{A21}},
  \bibinfo{pages}{3615} (\bibinfo{year}{2006}), \eprint{hep-ph/0510409}.

\bibitem[{\citenamefont{{de Paula} and Frederico}(2009)}]{dePaula:2009za}
\bibinfo{author}{\bibfnamefont{W.}~\bibnamefont{{de Paula}}} \bibnamefont{and}
  \bibinfo{author}{\bibfnamefont{T.}~\bibnamefont{Frederico}}
  (\bibinfo{year}{2009}), \eprint{0908.4282}.

\bibitem[{\citenamefont{Surovtsev et~al.}(2008)\citenamefont{Surovtsev,
  Bydzovsky, Kaminski, and Nagy}}]{Surovtsev:2008xr}
\bibinfo{author}{\bibfnamefont{Y.~S.} \bibnamefont{Surovtsev}},
  \bibinfo{author}{\bibfnamefont{P.}~\bibnamefont{Bydzovsky}},
  \bibinfo{author}{\bibfnamefont{R.}~\bibnamefont{Kaminski}}, \bibnamefont{and}
  \bibinfo{author}{\bibfnamefont{M.}~\bibnamefont{Nagy}}
  (\bibinfo{year}{2008}), \eprint{0811.0906}.

\bibitem[{\citenamefont{Cordon and Arriola}(2009)}]{Cordon:2009pj}
\bibinfo{author}{\bibfnamefont{A.~C.} \bibnamefont{Cordon}} \bibnamefont{and}
  \bibinfo{author}{\bibfnamefont{E.~R.} \bibnamefont{Arriola}}
  (\bibinfo{year}{2009}), \eprint{0905.4933}.

\bibitem[{\citenamefont{Colangelo et~al.}(2007)\citenamefont{Colangelo, {De
  Fazio}, Jugeau, and Nicotri}}]{Colangelo:2007pt}
\bibinfo{author}{\bibfnamefont{P.}~\bibnamefont{Colangelo}},
  \bibinfo{author}{\bibfnamefont{F.}~\bibnamefont{{De Fazio}}},
  \bibinfo{author}{\bibfnamefont{F.}~\bibnamefont{Jugeau}}, \bibnamefont{and}
  \bibinfo{author}{\bibfnamefont{S.}~\bibnamefont{Nicotri}},
  \bibinfo{journal}{Phys. Lett.} \textbf{\bibinfo{volume}{B652}},
  \bibinfo{pages}{73} (\bibinfo{year}{2007}), \eprint{hep-ph/0703316}.

\bibitem[{\citenamefont{Forkel}(2008)}]{Forkel:2007ru}
\bibinfo{author}{\bibfnamefont{H.}~\bibnamefont{Forkel}},
  \bibinfo{journal}{Phys. Rev.} \textbf{\bibinfo{volume}{D78}},
  \bibinfo{pages}{025001} (\bibinfo{year}{2008}), \eprint{0711.1179}.

\bibitem[{\citenamefont{Colangelo et~al.}(2008)\citenamefont{Colangelo, {De
  Fazio}, Giannuzzi, Jugeau, and Nicotri}}]{Colangelo:2008us}
\bibinfo{author}{\bibfnamefont{P.}~\bibnamefont{Colangelo}},
  \bibinfo{author}{\bibfnamefont{F.}~\bibnamefont{{De Fazio}}},
  \bibinfo{author}{\bibfnamefont{F.}~\bibnamefont{Giannuzzi}},
  \bibinfo{author}{\bibfnamefont{F.}~\bibnamefont{Jugeau}}, \bibnamefont{and}
  \bibinfo{author}{\bibfnamefont{S.}~\bibnamefont{Nicotri}},
  \bibinfo{journal}{Phys. Rev.} \textbf{\bibinfo{volume}{D78}},
  \bibinfo{pages}{055009} (\bibinfo{year}{2008}), \eprint{0807.1054}.

\bibitem[{\citenamefont{Afonin et~al.}(2008{\natexlab{a}})\citenamefont{Afonin,
  Andrianov, Andrianov, and Espriu}}]{Afonin:2008zz}
\bibinfo{author}{\bibfnamefont{S.~S.} \bibnamefont{Afonin}},
  \bibinfo{author}{\bibfnamefont{A.~A.} \bibnamefont{Andrianov}},
  \bibinfo{author}{\bibfnamefont{V.~A.} \bibnamefont{Andrianov}},
  \bibnamefont{and} \bibinfo{author}{\bibfnamefont{D.}~\bibnamefont{Espriu}},
  \bibinfo{journal}{AIP Conf. Proc.} \textbf{\bibinfo{volume}{1030}},
  \bibinfo{pages}{177} (\bibinfo{year}{2008}{\natexlab{a}}).

\bibitem[{\citenamefont{Zuo and Huang}(2008)}]{Zuo:2008re}
\bibinfo{author}{\bibfnamefont{F.}~\bibnamefont{Zuo}} \bibnamefont{and}
  \bibinfo{author}{\bibfnamefont{T.}~\bibnamefont{Huang}}
  (\bibinfo{year}{2008}), \eprint{0801.1172}.

\bibitem[{\citenamefont{Gherghetta et~al.}(2009)\citenamefont{Gherghetta,
  Kapusta, and Kelley}}]{Gherghetta:2009ac}
\bibinfo{author}{\bibfnamefont{T.}~\bibnamefont{Gherghetta}},
  \bibinfo{author}{\bibfnamefont{J.~I.} \bibnamefont{Kapusta}},
  \bibnamefont{and} \bibinfo{author}{\bibfnamefont{T.~M.}
  \bibnamefont{Kelley}}, \bibinfo{journal}{Phys. Rev.}
  \textbf{\bibinfo{volume}{D79}}, \bibinfo{pages}{076003}
  (\bibinfo{year}{2009}), \eprint{0902.1998}.

\bibitem[{\citenamefont{Afonin et~al.}(2008{\natexlab{b}})\citenamefont{Afonin,
  Andrianov, Andrianov, and Espriu}}]{Afonin:2008zza}
\bibinfo{author}{\bibfnamefont{S.~S.} \bibnamefont{Afonin}},
  \bibinfo{author}{\bibfnamefont{A.~A.} \bibnamefont{Andrianov}},
  \bibinfo{author}{\bibfnamefont{V.~A.} \bibnamefont{Andrianov}},
  \bibnamefont{and} \bibinfo{author}{\bibfnamefont{D.}~\bibnamefont{Espriu}},
  \bibinfo{journal}{AIP Conf. Proc.} \textbf{\bibinfo{volume}{1030}},
  \bibinfo{pages}{177} (\bibinfo{year}{2008}{\natexlab{b}}).

\bibitem[{\citenamefont{Collins et~al.}(1977)\citenamefont{Collins, Duncan, and
  Joglekar}}]{Collins:1976yq}
\bibinfo{author}{\bibfnamefont{J.~C.} \bibnamefont{Collins}},
  \bibinfo{author}{\bibfnamefont{A.}~\bibnamefont{Duncan}}, \bibnamefont{and}
  \bibinfo{author}{\bibfnamefont{S.~D.} \bibnamefont{Joglekar}},
  \bibinfo{journal}{Phys. Rev.} \textbf{\bibinfo{volume}{D16}},
  \bibinfo{pages}{438} (\bibinfo{year}{1977}).

\bibitem[{\citenamefont{Novikov et~al.}(1980)\citenamefont{Novikov, Shifman,
  Vainshtein, and Zakharov}}]{Novikov:1979va}
\bibinfo{author}{\bibfnamefont{V.~A.} \bibnamefont{Novikov}},
  \bibinfo{author}{\bibfnamefont{M.~A.} \bibnamefont{Shifman}},
  \bibinfo{author}{\bibfnamefont{A.~I.} \bibnamefont{Vainshtein}},
  \bibnamefont{and} \bibinfo{author}{\bibfnamefont{V.~I.}
  \bibnamefont{Zakharov}}, \bibinfo{journal}{Nucl. Phys.}
  \textbf{\bibinfo{volume}{B165}}, \bibinfo{pages}{67} (\bibinfo{year}{1980}).

\bibitem[{\citenamefont{Novikov et~al.}(1981)\citenamefont{Novikov, Shifman,
  Vainshtein, and Zakharov}}]{Novikov:1981xj}
\bibinfo{author}{\bibfnamefont{V.~A.} \bibnamefont{Novikov}},
  \bibinfo{author}{\bibfnamefont{M.~A.} \bibnamefont{Shifman}},
  \bibinfo{author}{\bibfnamefont{A.~I.} \bibnamefont{Vainshtein}},
  \bibnamefont{and} \bibinfo{author}{\bibfnamefont{V.~I.}
  \bibnamefont{Zakharov}}, \bibinfo{journal}{Nucl. Phys.}
  \textbf{\bibinfo{volume}{B191}}, \bibinfo{pages}{301} (\bibinfo{year}{1981}).

\bibitem[{\citenamefont{Pascual and Tarrach}(1982)}]{Pascual:1982bv}
\bibinfo{author}{\bibfnamefont{P.}~\bibnamefont{Pascual}} \bibnamefont{and}
  \bibinfo{author}{\bibfnamefont{R.}~\bibnamefont{Tarrach}},
  \bibinfo{journal}{Phys. Lett.} \textbf{\bibinfo{volume}{B113}},
  \bibinfo{pages}{495} (\bibinfo{year}{1982}).

\bibitem[{\citenamefont{Dominguez and Paver}(1986)}]{Dominguez:1986td}
\bibinfo{author}{\bibfnamefont{C.~A.} \bibnamefont{Dominguez}}
  \bibnamefont{and} \bibinfo{author}{\bibfnamefont{N.}~\bibnamefont{Paver}},
  \bibinfo{journal}{Z. Phys.} \textbf{\bibinfo{volume}{C31}},
  \bibinfo{pages}{591} (\bibinfo{year}{1986}).

\bibitem[{\citenamefont{Narison}(1998)}]{Narison:1996fm}
\bibinfo{author}{\bibfnamefont{S.}~\bibnamefont{Narison}},
  \bibinfo{journal}{Nucl. Phys.} \textbf{\bibinfo{volume}{B509}},
  \bibinfo{pages}{312} (\bibinfo{year}{1998}), \eprint{hep-ph/9612457}.

\bibitem[{\citenamefont{Donoghue and Leutwyler}(1991)}]{Donoghue:1991qv}
\bibinfo{author}{\bibfnamefont{J.~F.} \bibnamefont{Donoghue}} \bibnamefont{and}
  \bibinfo{author}{\bibfnamefont{H.}~\bibnamefont{Leutwyler}},
  \bibinfo{journal}{Z. Phys.} \textbf{\bibinfo{volume}{C52}},
  \bibinfo{pages}{343} (\bibinfo{year}{1991}).

\bibitem[{\citenamefont{Narison and Veneziano}(1989)}]{Narison:1988ts}
\bibinfo{author}{\bibfnamefont{S.}~\bibnamefont{Narison}} \bibnamefont{and}
  \bibinfo{author}{\bibfnamefont{G.}~\bibnamefont{Veneziano}},
  \bibinfo{journal}{Int. J. Mod. Phys.} \textbf{\bibinfo{volume}{A4}},
  \bibinfo{pages}{2751} (\bibinfo{year}{1989}).

\bibitem[{\citenamefont{Callan et~al.}(1970)\citenamefont{Callan, Coleman, and
  Jackiw}}]{Callan:1970ze}
\bibinfo{author}{\bibfnamefont{C.~G.~J.} \bibnamefont{Callan}},
  \bibinfo{author}{\bibfnamefont{S.~R.} \bibnamefont{Coleman}},
  \bibnamefont{and} \bibinfo{author}{\bibfnamefont{R.}~\bibnamefont{Jackiw}},
  \bibinfo{journal}{Ann. Phys.} \textbf{\bibinfo{volume}{59}},
  \bibinfo{pages}{42} (\bibinfo{year}{1970}).

\bibitem[{\citenamefont{Schechter}(1980)}]{Schechter:1980ak}
\bibinfo{author}{\bibfnamefont{J.}~\bibnamefont{Schechter}},
  \bibinfo{journal}{Phys. Rev.} \textbf{\bibinfo{volume}{D21}},
  \bibinfo{pages}{3393} (\bibinfo{year}{1980}).

\bibitem[{\citenamefont{Migdal and Shifman}(1982)}]{Migdal:1982jp}
\bibinfo{author}{\bibfnamefont{A.~A.} \bibnamefont{Migdal}} \bibnamefont{and}
  \bibinfo{author}{\bibfnamefont{M.~A.} \bibnamefont{Shifman}},
  \bibinfo{journal}{Phys. Lett.} \textbf{\bibinfo{volume}{B114}},
  \bibinfo{pages}{445} (\bibinfo{year}{1982}).

\bibitem[{\citenamefont{Gomm et~al.}(1986)\citenamefont{Gomm, Jain, Johnson,
  and Schechter}}]{Gomm:1985ut}
\bibinfo{author}{\bibfnamefont{R.}~\bibnamefont{Gomm}},
  \bibinfo{author}{\bibfnamefont{P.}~\bibnamefont{Jain}},
  \bibinfo{author}{\bibfnamefont{R.}~\bibnamefont{Johnson}}, \bibnamefont{and}
  \bibinfo{author}{\bibfnamefont{J.}~\bibnamefont{Schechter}},
  \bibinfo{journal}{Phys. Rev.} \textbf{\bibinfo{volume}{D33}},
  \bibinfo{pages}{801} (\bibinfo{year}{1986}).

\bibitem[{\citenamefont{Ellis and Lanik}(1985)}]{Ellis:1984jv}
\bibinfo{author}{\bibfnamefont{J.~R.} \bibnamefont{Ellis}} \bibnamefont{and}
  \bibinfo{author}{\bibfnamefont{J.}~\bibnamefont{Lanik}},
  \bibinfo{journal}{Phys. Lett.} \textbf{\bibinfo{volume}{B150}},
  \bibinfo{pages}{289} (\bibinfo{year}{1985}).

\bibitem[{\citenamefont{Carruthers}(1971)}]{Carruthers:1971vz}
\bibinfo{author}{\bibfnamefont{P.}~\bibnamefont{Carruthers}},
  \bibinfo{journal}{Phys. Rept.} \textbf{\bibinfo{volume}{1}},
  \bibinfo{pages}{1} (\bibinfo{year}{1971}).

\bibitem[{\citenamefont{Brommel et~al.}(2007)}]{Brommel:2007xd}
\bibinfo{author}{\bibfnamefont{D.}~\bibnamefont{Brommel}} \bibnamefont{et~al.}
  (\bibinfo{collaboration}{QCDSF}) (\bibinfo{year}{2007}),
  \eprint{arXiv:0708.2249 [hep-lat]}.

\bibitem[{\citenamefont{Brommel}(2007)}]{Brommel:PhD}
\bibinfo{author}{\bibfnamefont{D.}~\bibnamefont{Brommel}}, Ph.D. thesis,
  \bibinfo{school}{{University of Regensburg}}, \bibinfo{address}{{Regensburg,
  Germany}} (\bibinfo{year}{2007}), \bibinfo{note}{dESY-THESIS-2007-023}.

\bibitem[{\citenamefont{Broniowski and Arriola}(2009)}]{Broniowski:2009zh}
\bibinfo{author}{\bibfnamefont{W.}~\bibnamefont{Broniowski}} \bibnamefont{and}
  \bibinfo{author}{\bibfnamefont{E.~R.} \bibnamefont{Arriola}},
  \bibinfo{journal}{Phys. Rev.} \textbf{\bibinfo{volume}{D79}},
  \bibinfo{pages}{057501} (\bibinfo{year}{2009}), \eprint{0901.3336}.

\bibitem[{\citenamefont{Sutton et~al.}(1992)\citenamefont{Sutton, Martin,
  Roberts, and Stirling}}]{Sutton:1991ay}
\bibinfo{author}{\bibfnamefont{P.~J.} \bibnamefont{Sutton}},
  \bibinfo{author}{\bibfnamefont{A.~D.} \bibnamefont{Martin}},
  \bibinfo{author}{\bibfnamefont{R.~G.} \bibnamefont{Roberts}},
  \bibnamefont{and} \bibinfo{author}{\bibfnamefont{W.~J.}
  \bibnamefont{Stirling}}, \bibinfo{journal}{Phys. Rev.}
  \textbf{\bibinfo{volume}{D45}}, \bibinfo{pages}{2349} (\bibinfo{year}{1992}).

\bibitem[{\citenamefont{Conway et~al.}(1989)}]{Conway:1989fs}
\bibinfo{author}{\bibfnamefont{J.~S.} \bibnamefont{Conway}}
  \bibnamefont{et~al.}, \bibinfo{journal}{Phys. Rev.}
  \textbf{\bibinfo{volume}{D39}}, \bibinfo{pages}{92} (\bibinfo{year}{1989}).

\bibitem[{\citenamefont{Gluck et~al.}(1999)\citenamefont{Gluck, Reya, and
  Schienbein}}]{Gluck:1999xe}
\bibinfo{author}{\bibfnamefont{M.}~\bibnamefont{Gluck}},
  \bibinfo{author}{\bibfnamefont{E.}~\bibnamefont{Reya}}, \bibnamefont{and}
  \bibinfo{author}{\bibfnamefont{I.}~\bibnamefont{Schienbein}},
  \bibinfo{journal}{Eur. Phys. J.} \textbf{\bibinfo{volume}{C10}},
  \bibinfo{pages}{313} (\bibinfo{year}{1999}), \eprint{hep-ph/9903288}.

\bibitem[{\citenamefont{Broniowski and Arriola}(2008)}]{Broniowski:2008hx}
\bibinfo{author}{\bibfnamefont{W.}~\bibnamefont{Broniowski}} \bibnamefont{and}
  \bibinfo{author}{\bibfnamefont{E.~R.} \bibnamefont{Arriola}},
  \bibinfo{journal}{Phys. Rev.} \textbf{\bibinfo{volume}{D78}},
  \bibinfo{pages}{094011} (\bibinfo{year}{2008}), \eprint{0809.1744}.

\bibitem[{\citenamefont{Dominguez}(2001)}]{Dominguez:2001zu}
\bibinfo{author}{\bibfnamefont{C.~A.} \bibnamefont{Dominguez}},
  \bibinfo{journal}{Phys. Lett.} \textbf{\bibinfo{volume}{B512}},
  \bibinfo{pages}{331} (\bibinfo{year}{2001}), \eprint{hep-ph/0102190}.

\bibitem[{\citenamefont{Ruiz~Arriola and
  Broniowski}(2008)}]{RuizArriola:2008sq}
\bibinfo{author}{\bibfnamefont{E.}~\bibnamefont{Ruiz~Arriola}}
  \bibnamefont{and}
  \bibinfo{author}{\bibfnamefont{W.}~\bibnamefont{Broniowski}},
  \bibinfo{journal}{Phys. Rev.} \textbf{\bibinfo{volume}{D78}},
  \bibinfo{pages}{034031} (\bibinfo{year}{2008}), \eprint{0807.3488}.

\bibitem[{\citenamefont{Gockeler et~al.}(2004)}]{Gockeler:2003jfa}
\bibinfo{author}{\bibfnamefont{M.}~\bibnamefont{Gockeler}} \bibnamefont{et~al.}
  (\bibinfo{collaboration}{QCDSF}), \bibinfo{journal}{Phys. Rev. Lett.}
  \textbf{\bibinfo{volume}{92}}, \bibinfo{pages}{042002}
  (\bibinfo{year}{2004}), \eprint{hep-ph/0304249}.

\bibitem[{\citenamefont{Hagler et~al.}(2008)}]{Hagler:2007xi}
\bibinfo{author}{\bibfnamefont{P.}~\bibnamefont{Hagler}} \bibnamefont{et~al.}
  (\bibinfo{collaboration}{LHPC}), \bibinfo{journal}{Phys. Rev.}
  \textbf{\bibinfo{volume}{D77}}, \bibinfo{pages}{094502}
  (\bibinfo{year}{2008}), \eprint{0705.4295}.

\bibitem[{\citenamefont{Ioffe and Zyablyuk}(2003)}]{Ioffe:2002be}
\bibinfo{author}{\bibfnamefont{B.~L.} \bibnamefont{Ioffe}} \bibnamefont{and}
  \bibinfo{author}{\bibfnamefont{K.~N.} \bibnamefont{Zyablyuk}},
  \bibinfo{journal}{Eur. Phys. J.} \textbf{\bibinfo{volume}{C27}},
  \bibinfo{pages}{229} (\bibinfo{year}{2003}), \eprint{hep-ph/0207183}.

\bibitem[{\citenamefont{Golterman and Peris}(2001)}]{Golterman:2001nk}
\bibinfo{author}{\bibfnamefont{M.}~\bibnamefont{Golterman}} \bibnamefont{and}
  \bibinfo{author}{\bibfnamefont{S.}~\bibnamefont{Peris}},
  \bibinfo{journal}{JHEP} \textbf{\bibinfo{volume}{01}}, \bibinfo{pages}{028}
  (\bibinfo{year}{2001}), \eprint{hep-ph/0101098}.

\bibitem[{\citenamefont{Beane}(2001{\natexlab{a}})}]{Beane:2001uj}
\bibinfo{author}{\bibfnamefont{S.~R.} \bibnamefont{Beane}},
  \bibinfo{journal}{Phys. Rev.} \textbf{\bibinfo{volume}{D64}},
  \bibinfo{pages}{116010} (\bibinfo{year}{2001}{\natexlab{a}}),
  \eprint{hep-ph/0106022}.

\bibitem[{\citenamefont{Beane}(2001{\natexlab{b}})}]{Beane:2001em}
\bibinfo{author}{\bibfnamefont{S.~R.} \bibnamefont{Beane}},
  \bibinfo{journal}{Phys. Lett.} \textbf{\bibinfo{volume}{B521}},
  \bibinfo{pages}{47} (\bibinfo{year}{2001}{\natexlab{b}}),
  \eprint{hep-ph/0108025}.

\bibitem[{\citenamefont{Simonov}(2002)}]{Simonov:2001di}
\bibinfo{author}{\bibfnamefont{Y.~A.} \bibnamefont{Simonov}},
  \bibinfo{journal}{Phys. Atom. Nucl.} \textbf{\bibinfo{volume}{65}},
  \bibinfo{pages}{135} (\bibinfo{year}{2002}), \eprint{hep-ph/0109081}.

\bibitem[{\citenamefont{Golterman and Peris}(2003)}]{Golterman:2002mi}
\bibinfo{author}{\bibfnamefont{M.}~\bibnamefont{Golterman}} \bibnamefont{and}
  \bibinfo{author}{\bibfnamefont{S.}~\bibnamefont{Peris}},
  \bibinfo{journal}{Phys. Rev.} \textbf{\bibinfo{volume}{D67}},
  \bibinfo{pages}{096001} (\bibinfo{year}{2003}), \eprint{hep-ph/0207060}.

\bibitem[{\citenamefont{Afonin}(2003)}]{Afonin:2003gp}
\bibinfo{author}{\bibfnamefont{S.~S.} \bibnamefont{Afonin}},
  \bibinfo{journal}{Phys. Lett.} \textbf{\bibinfo{volume}{B576}},
  \bibinfo{pages}{122} (\bibinfo{year}{2003}), \eprint{hep-ph/0309337}.

\bibitem[{\citenamefont{Afonin et~al.}(2004)\citenamefont{Afonin, Andrianov,
  Andrianov, and Espriu}}]{Afonin:2004yb}
\bibinfo{author}{\bibfnamefont{S.~S.} \bibnamefont{Afonin}},
  \bibinfo{author}{\bibfnamefont{A.~A.} \bibnamefont{Andrianov}},
  \bibinfo{author}{\bibfnamefont{V.~A.} \bibnamefont{Andrianov}},
  \bibnamefont{and} \bibinfo{author}{\bibfnamefont{D.}~\bibnamefont{Espriu}},
  \bibinfo{journal}{JHEP} \textbf{\bibinfo{volume}{04}}, \bibinfo{pages}{039}
  (\bibinfo{year}{2004}), \eprint{hep-ph/0403268}.

\bibitem[{\citenamefont{{Ruiz Arriola} and
  Broniowski}(2006)}]{RuizArriola:2006gq}
\bibinfo{author}{\bibfnamefont{E.}~\bibnamefont{{Ruiz Arriola}}}
  \bibnamefont{and}
  \bibinfo{author}{\bibfnamefont{W.}~\bibnamefont{Broniowski}},
  \bibinfo{journal}{Phys. Rev.} \textbf{\bibinfo{volume}{D73}},
  \bibinfo{pages}{097502} (\bibinfo{year}{2006}), \eprint{hep-ph/0603263}.

\bibitem[{\citenamefont{Ruiz~Arriola and Broniowski}(2006)}]{Arriola:2006ii}
\bibinfo{author}{\bibfnamefont{E.}~\bibnamefont{Ruiz~Arriola}}
  \bibnamefont{and}
  \bibinfo{author}{\bibfnamefont{W.}~\bibnamefont{Broniowski}},
  \bibinfo{journal}{Phys. Rev.} \textbf{\bibinfo{volume}{D74}},
  \bibinfo{pages}{034008} (\bibinfo{year}{2006}), \eprint{hep-ph/0605318}.

\bibitem[{\citenamefont{Arriola and Broniowski}(2007)}]{Arriola:2006sv}
\bibinfo{author}{\bibfnamefont{E.~R.} \bibnamefont{Arriola}} \bibnamefont{and}
  \bibinfo{author}{\bibfnamefont{W.}~\bibnamefont{Broniowski}},
  \bibinfo{journal}{Eur. Phys. J.} \textbf{\bibinfo{volume}{A31}},
  \bibinfo{pages}{739} (\bibinfo{year}{2007}), \eprint{hep-ph/0609266}.

\bibitem[{\citenamefont{Afonin}(2007)}]{Afonin:2007mj}
\bibinfo{author}{\bibfnamefont{S.~S.} \bibnamefont{Afonin}},
  \bibinfo{journal}{Int. J. Mod. Phys.} \textbf{\bibinfo{volume}{A22}},
  \bibinfo{pages}{4537} (\bibinfo{year}{2007}), \eprint{0704.1639}.

\bibitem[{\citenamefont{Celenza and Shakin}(1986)}]{Celenza:1986th}
\bibinfo{author}{\bibfnamefont{L.~S.} \bibnamefont{Celenza}} \bibnamefont{and}
  \bibinfo{author}{\bibfnamefont{C.~M.} \bibnamefont{Shakin}},
  \bibinfo{journal}{Phys. Rev.} \textbf{\bibinfo{volume}{D34}},
  \bibinfo{pages}{1591} (\bibinfo{year}{1986}).

\bibitem[{\citenamefont{Zakharov}(2005)}]{Zakharov:2005cg}
\bibinfo{author}{\bibfnamefont{V.~I.} \bibnamefont{Zakharov}}
  (\bibinfo{year}{2005}), \eprint{hep-ph/0509114}.

\bibitem[{\citenamefont{Narison}(2005)}]{Narison:2005hb}
\bibinfo{author}{\bibfnamefont{S.}~\bibnamefont{Narison}}
  (\bibinfo{year}{2005}), \eprint{hep-ph/0508259}.

\bibitem[{\citenamefont{Hutter}(1993)}]{Hutter:1993sc}
\bibinfo{author}{\bibfnamefont{M.}~\bibnamefont{Hutter}}
  (\bibinfo{year}{1993}), \eprint{hep-ph/9501335}.

\bibitem[{\citenamefont{Chetyrkin et~al.}(1999)\citenamefont{Chetyrkin,
  Narison, and Zakharov}}]{Chetyrkin:1998yr}
\bibinfo{author}{\bibfnamefont{K.~G.} \bibnamefont{Chetyrkin}},
  \bibinfo{author}{\bibfnamefont{S.}~\bibnamefont{Narison}}, \bibnamefont{and}
  \bibinfo{author}{\bibfnamefont{V.~I.} \bibnamefont{Zakharov}},
  \bibinfo{journal}{Nucl. Phys.} \textbf{\bibinfo{volume}{B550}},
  \bibinfo{pages}{353} (\bibinfo{year}{1999}), \eprint{hep-ph/9811275}.

\bibitem[{\citenamefont{Dominguez}(1995)}]{Dominguez:1994qt}
\bibinfo{author}{\bibfnamefont{C.~A.} \bibnamefont{Dominguez}},
  \bibinfo{journal}{Phys. Lett.} \textbf{\bibinfo{volume}{B345}},
  \bibinfo{pages}{291} (\bibinfo{year}{1995}), \eprint{hep-ph/9411331}.

\bibitem[{\citenamefont{Gubarev et~al.}(2001)\citenamefont{Gubarev, Stodolsky,
  and Zakharov}}]{Gubarev:2000eu}
\bibinfo{author}{\bibfnamefont{F.~V.} \bibnamefont{Gubarev}},
  \bibinfo{author}{\bibfnamefont{L.}~\bibnamefont{Stodolsky}},
  \bibnamefont{and} \bibinfo{author}{\bibfnamefont{V.~I.}
  \bibnamefont{Zakharov}}, \bibinfo{journal}{Phys. Rev. Lett.}
  \textbf{\bibinfo{volume}{86}}, \bibinfo{pages}{2220} (\bibinfo{year}{2001}),
  \eprint{hep-ph/0010057}.

\bibitem[{\citenamefont{Gubarev and Zakharov}(2001)}]{Gubarev:2000nz}
\bibinfo{author}{\bibfnamefont{F.~V.} \bibnamefont{Gubarev}} \bibnamefont{and}
  \bibinfo{author}{\bibfnamefont{V.~I.} \bibnamefont{Zakharov}},
  \bibinfo{journal}{Phys. Lett.} \textbf{\bibinfo{volume}{B501}},
  \bibinfo{pages}{28} (\bibinfo{year}{2001}), \eprint{hep-ph/0010096}.

\bibitem[{\citenamefont{Kondo}(2001)}]{Kondo:2001nq}
\bibinfo{author}{\bibfnamefont{K.-I.} \bibnamefont{Kondo}},
  \bibinfo{journal}{Phys. Lett.} \textbf{\bibinfo{volume}{B514}},
  \bibinfo{pages}{335} (\bibinfo{year}{2001}), \eprint{hep-th/0105299}.

\bibitem[{\citenamefont{Verschelde et~al.}(2001)\citenamefont{Verschelde,
  Knecht, Van~Acoleyen, and Vanderkelen}}]{Verschelde:2001ia}
\bibinfo{author}{\bibfnamefont{H.}~\bibnamefont{Verschelde}},
  \bibinfo{author}{\bibfnamefont{K.}~\bibnamefont{Knecht}},
  \bibinfo{author}{\bibfnamefont{K.}~\bibnamefont{Van~Acoleyen}},
  \bibnamefont{and}
  \bibinfo{author}{\bibfnamefont{M.}~\bibnamefont{Vanderkelen}},
  \bibinfo{journal}{Phys. Lett.} \textbf{\bibinfo{volume}{B516}},
  \bibinfo{pages}{307} (\bibinfo{year}{2001}), \eprint{hep-th/0105018}.

\bibitem[{\citenamefont{Dorokhov and Broniowski}(2003)}]{Dorokhov:2003kf}
\bibinfo{author}{\bibfnamefont{A.~E.} \bibnamefont{Dorokhov}} \bibnamefont{and}
  \bibinfo{author}{\bibfnamefont{W.}~\bibnamefont{Broniowski}},
  \bibinfo{journal}{Eur. Phys. J.} \textbf{\bibinfo{volume}{C32}},
  \bibinfo{pages}{79} (\bibinfo{year}{2003}), \eprint{hep-ph/0305037}.

\bibitem[{\citenamefont{Dorokhov}(2006)}]{Dorokhov:2006ac}
\bibinfo{author}{\bibfnamefont{A.~E.} \bibnamefont{Dorokhov}}
  (\bibinfo{year}{2006}), \eprint{hep-ph/0601114}.

\bibitem[{\citenamefont{Boucaud et~al.}(2001)}]{Boucaud:2001st}
\bibinfo{author}{\bibfnamefont{P.}~\bibnamefont{Boucaud}} \bibnamefont{et~al.},
  \bibinfo{journal}{Phys. Rev.} \textbf{\bibinfo{volume}{D63}},
  \bibinfo{pages}{114003} (\bibinfo{year}{2001}), \eprint{hep-ph/0101302}.

\bibitem[{\citenamefont{Ruiz~Arriola et~al.}(2004)\citenamefont{Ruiz~Arriola,
  Bowman, and Broniowski}}]{RuizArriola:2004en}
\bibinfo{author}{\bibfnamefont{E.}~\bibnamefont{Ruiz~Arriola}},
  \bibinfo{author}{\bibfnamefont{P.~O.} \bibnamefont{Bowman}},
  \bibnamefont{and}
  \bibinfo{author}{\bibfnamefont{W.}~\bibnamefont{Broniowski}},
  \bibinfo{journal}{Phys. Rev.} \textbf{\bibinfo{volume}{D70}},
  \bibinfo{pages}{097505} (\bibinfo{year}{2004}), \eprint{hep-ph/0408309}.

\bibitem[{\citenamefont{Megias et~al.}(2006)\citenamefont{Megias, Ruiz~Arriola,
  and Salcedo}}]{Megias:2005ve}
\bibinfo{author}{\bibfnamefont{E.}~\bibnamefont{Megias}},
  \bibinfo{author}{\bibfnamefont{E.}~\bibnamefont{Ruiz~Arriola}},
  \bibnamefont{and} \bibinfo{author}{\bibfnamefont{L.~L.}
  \bibnamefont{Salcedo}}, \bibinfo{journal}{JHEP}
  \textbf{\bibinfo{volume}{01}}, \bibinfo{pages}{073} (\bibinfo{year}{2006}),
  \eprint{hep-ph/0505215}.

\bibitem[{\citenamefont{Narison}(1993)}]{Narison:1992ru}
\bibinfo{author}{\bibfnamefont{S.}~\bibnamefont{Narison}},
  \bibinfo{journal}{Phys. Lett.} \textbf{\bibinfo{volume}{B300}},
  \bibinfo{pages}{293} (\bibinfo{year}{1993}).

\bibitem[{\citenamefont{Dominguez and Schilcher}(2000)}]{Dominguez:1999xa}
\bibinfo{author}{\bibfnamefont{C.~A.} \bibnamefont{Dominguez}}
  \bibnamefont{and}
  \bibinfo{author}{\bibfnamefont{K.}~\bibnamefont{Schilcher}},
  \bibinfo{journal}{Phys. Rev.} \textbf{\bibinfo{volume}{D61}},
  \bibinfo{pages}{114020} (\bibinfo{year}{2000}), \eprint{hep-ph/9903483}.

\bibitem[{\citenamefont{Dominguez and Schilcher}(2007)}]{Dominguez:2006ct}
\bibinfo{author}{\bibfnamefont{C.~A.} \bibnamefont{Dominguez}}
  \bibnamefont{and}
  \bibinfo{author}{\bibfnamefont{K.}~\bibnamefont{Schilcher}},
  \bibinfo{journal}{JHEP} \textbf{\bibinfo{volume}{01}}, \bibinfo{pages}{093}
  (\bibinfo{year}{2007}), \eprint{hep-ph/0611347}.

\bibitem[{\citenamefont{Shifman}(2000)}]{Shifman:2000jv}
\bibinfo{author}{\bibfnamefont{M.~A.} \bibnamefont{Shifman}}
  (\bibinfo{year}{2000}), \eprint{hep-ph/0009131}.

\bibitem[{\citenamefont{Narison and Zakharov}(2009)}]{Narison:2009ag}
\bibinfo{author}{\bibfnamefont{S.}~\bibnamefont{Narison}} \bibnamefont{and}
  \bibinfo{author}{\bibfnamefont{V.~I.} \bibnamefont{Zakharov}},
  \bibinfo{journal}{Phys. Lett.} \textbf{\bibinfo{volume}{B679}},
  \bibinfo{pages}{355} (\bibinfo{year}{2009}), \eprint{0906.4312}.

\bibitem[{\citenamefont{Megias et~al.}(2009)\citenamefont{Megias, Arriola, and
  Salcedo}}]{Megias:2009ar}
\bibinfo{author}{\bibfnamefont{E.}~\bibnamefont{Megias}},
  \bibinfo{author}{\bibfnamefont{E.~R.} \bibnamefont{Arriola}},
  \bibnamefont{and} \bibinfo{author}{\bibfnamefont{L.~L.}
  \bibnamefont{Salcedo}} (\bibinfo{year}{2009}), \eprint{0912.0499}.

\bibitem[{\citenamefont{Narison}(1996)}]{Narison:1995tw}
\bibinfo{author}{\bibfnamefont{S.}~\bibnamefont{Narison}},
  \bibinfo{journal}{Phys. Lett.} \textbf{\bibinfo{volume}{B387}},
  \bibinfo{pages}{162} (\bibinfo{year}{1996}), \eprint{hep-ph/9512348}.

\bibitem[{\citenamefont{Davier et~al.}(2008)\citenamefont{Davier,
  Descotes-Genon, Hocker, Malaescu, and Zhang}}]{Davier:2008sk}
\bibinfo{author}{\bibfnamefont{M.}~\bibnamefont{Davier}},
  \bibinfo{author}{\bibfnamefont{S.}~\bibnamefont{Descotes-Genon}},
  \bibinfo{author}{\bibfnamefont{A.}~\bibnamefont{Hocker}},
  \bibinfo{author}{\bibfnamefont{B.}~\bibnamefont{Malaescu}}, \bibnamefont{and}
  \bibinfo{author}{\bibfnamefont{Z.}~\bibnamefont{Zhang}},
  \bibinfo{journal}{Eur. Phys. J.} \textbf{\bibinfo{volume}{C56}},
  \bibinfo{pages}{305} (\bibinfo{year}{2008}), \eprint{0803.0979}.

\bibitem[{\citenamefont{Broniowski and Golli}(2003)}]{Broniowski:2002ew}
\bibinfo{author}{\bibfnamefont{W.}~\bibnamefont{Broniowski}} \bibnamefont{and}
  \bibinfo{author}{\bibfnamefont{B.}~\bibnamefont{Golli}},
  \bibinfo{journal}{Nucl. Phys.} \textbf{\bibinfo{volume}{A714}},
  \bibinfo{pages}{575} (\bibinfo{year}{2003}), \eprint{hep-ph/0210200}.

\bibitem[{\citenamefont{Beane and van Kolck}(1994)}]{Beane:1994ds}
\bibinfo{author}{\bibfnamefont{S.~R.} \bibnamefont{Beane}} \bibnamefont{and}
  \bibinfo{author}{\bibfnamefont{U.}~\bibnamefont{van Kolck}},
  \bibinfo{journal}{Phys. Lett.} \textbf{\bibinfo{volume}{B328}},
  \bibinfo{pages}{137} (\bibinfo{year}{1994}), \eprint{hep-ph/9401218}.

\end{thebibliography}

\end{document}